\documentstyle[11pt,fleqn]{article}     
\def\etal{{\it et al.\thinspace}}

\def\cf{{\it c.f.\ }}
\def\nH{\hbox{$N_{\rm H}$}}

\def\h50{\hbox{$h_{50}$\,}}
\def\tpow#1{$10^{#1}\,$}            
\def\spose#1{\hbox to 0pt{#1\hss}}
\def\ltsimm{\mathrel{\spose{\lower 3pt\hbox{$\sim$}}
    \raise 2.0pt\hbox{$<$}}}
\def\ltsim{$\mathrel{\spose{\lower 3pt\hbox{$\sim$}}
    \raise 2.0pt\hbox{$<$}}$}
\def\gtsimm{\mathrel{\spose{\lower 3pt\hbox{$\sim$}}
    \raise 2.0pt\hbox{$>$}}}
\def\gtsim{$\mathrel{\spose{\lower 3pt\hbox{$\sim$}}
    \raise 2.0pt\hbox{$>$}}$}

\def\fract#1/#2{\leavevmode\kern.1em                   
   \raise.5ex\hbox{\the\scriptfont0 #1}\kern-.1em
   /\kern-.15em\lower.25ex\hbox{\the\scriptfont0 #2}}
\def\deg{\hbox{$^\circ$}}
\def\arcm{\hbox{$^\prime$}}

\def\cm{{\rm\thinspace cm}}
\def\erg{{\rm\thinspace erg}}

\def\keV{{\rm\thinspace keV}}
\def\km{{\rm\thinspace km}}
\def\kpc{{\rm\thinspace kpc}}
\def\Lsol{\hbox{$\thinspace L_{\odot}$}}

\def\Mpc{{\rm\thinspace Mpc}}
\def\Msol{\hbox{$\thinspace M_{\odot}$}}
\def\pc{{\rm\thinspace pc}}
\def\s{{\rm\thinspace s}}

\def\pcc{\hbox{$\cm^{-3}\,$}}

\def\pcm2{\hbox{$\cm^{-2}\,$}}
\def\ergpcm3ps{\hbox{$\erg\cm^{-3}\s^{-1}\,$}}
\def\ergps{\hbox{$\erg\s^{-1}\,$}}

\def\kmps{\hbox{$\km\s^{-1}\,$}}
\def\kmpspMpc{\hbox{$\km\s^{-1}\Mpc^{-1}\,$}}
\def\Lsolppc3{\hbox{$\Lsol\pc^{-3}\,$}}
\def\Msolppc3{\hbox{$\Msol\pc^{-3}\,$}}

\def\hi{H{\sc i}}

\def\ROS{{\it ROSAT}}
\def\fsp{$f_{\rm sp}$}

\begin{document}                            

\begin{center}
{\Large \bf A \ROS\ survey of Hickson's compact galaxy groups}\\[10mm]
\end{center}

{\large T.J. Ponman$^1$, P.D.J. Bourner$^1$, H. Ebeling$^2$ and
H. B\"ohringer$^3$}\\[1mm]
$^1$School of Physics and Space
Research, University of Birmingham, Edgbaston, Birmingham B15 2TT, UK \\
$^2$Institute of Astronomy, Madingley Road, Cambridge CB3 0HA, UK\\
$^3$Max-Planck-Intitut f\"ur Extraterrestrische Physik, D-85740 Garching,
Germany\\[8mm]
\section*{ABSTRACT}
We report the results of an almost complete survey of the X-ray
properties of Hickson's compact galaxy groups with the \ROS\ PSPC.
Diffuse X-ray emission is detected from 22 groups. We infer that
hot intragroup gas is present in $\gtsimm 75\%$ of these systems
and derive their X-ray luminosity function.
Earlier reports that only spiral-poor systems exhibit diffuse X-ray
emission are found to be incorrect.
Strong correlations are found between the X-ray luminosity and both the
gas temperature and the velocity dispersion of the group galaxies.
We argue that these properties provide strong evidence that most
of these groups are genuinely compact configurations, rather than
line-of-sight superpositions. Comparison with the X-ray properties
of galaxy clusters indicate a significant steepening of the $L:T$
relation below $T\sim1$\keV, which may result from the action of galaxy winds.

\section{INTRODUCTION}
\label{sec:intro}
Approximately half of all galaxies are found within groups which are
probably bound (Tully 1987). Whilst groups are very numerous, they are
hard to identify with confidence as gravitationally bound systems, even
in redshift catalogues, due to the problems of contamination and small
number statistics which attend such poor systems.  {\it Compact galaxy
groups} are ones in which the galaxies are separated on the sky by only a
few galactic radii. Only a small fraction of groups fall into this
category, but due to their sometimes spectacular appearance, and to the
comparative ease with which they can be discriminated from the field,
these systems have received a good deal of attention.

The most widely studied collection of compact groups are the Hickson
compact groups (HCGs) compiled by Hickson (1982) by applying a
set of well defined criteria for population, surface density and
isolation to the Palomar Sky Survey (E band) plate collection.  Numerous
follow-up studies of Hickson's catalogue of 100 groups have shown that
many galaxies in HCGs show morphological or kinematical peculiarities
(Rubin, Hunter \& Ford 1991; Mendes de Oliveira \& Hickson 1994), that,
relative to field galaxies, spirals tend to have low HI content (Williams
\& Rood 1987), and that
ellipticals have low velocity dispersion for their luminosity (Zepf \&
Whitmore 1993). Hickson, Kindl \& Huchra (1988) have also noted the
existence of statistical correlations in group properties, such as
a tendency for low spiral fraction to be associated with high velocity
dispersion.

Despite this intensive observational study, and increasing theoretical
interest, the very nature of compact groups is still unclear.  Numerical
simulations and theoretical arguments (e.g. Barnes 1984, 1985; Mamon
1987; Barnes 1989) indicate that, at the high densities inferred in
compact groups, galaxies should merge within a few crossing times.  In
practice, the way in which compact groups are identified (i.e. through
their high
projected galaxy density) imposes strong selection effects which are likely to
cause dynamical timescales to be underestimated (White 1990). However, these
effects are hard to quantify, and Mamon (1986, 1987, 1995) has argued for
several years that the difficulties of forming dense bound groups at the
rate required to replace those lost through merging make it much more
likely that the bulk of them are not truly dense in three dimensions. The
observed peculiarities in the properties of HCG galaxies may still be
explained if they contain real pairs, or occasionally triplets, which can
interact (Mamon 1992a), whilst trends such as the relationship between
spiral fraction and velocity dispersion could result from
redshift-dependent selection effects (Mamon 1992b).  A new twist has
recently been added to the `chance superposition' hypothesis by the
suggestion of Hernquist, Katz \& Weinberg (1995) (HKW95), that compact
groups may be the result of looking along the filamentary arrangements of
galaxies which are predicted by most cosmological simulations.

Meanwhile, recent numerical studies have suggested two ways in which
compact groups could be continuously generated to replace those lost to
the merging instability. Diaferio, Geller \& Ramella (1994, 1995)
(hereafter DGR94, DGR95) argue that compact configurations, most of which
are bound, are formed continuously during the collapse and virialisation
of larger groups -- though it should be noted (Mamon, private
communication) that they do not include the requirement of high projected
galaxy density ($<26^{\rm m}$arcsec$^{-2}$) applied by Hickson, when
identifying `compact groups' in their simulated data. Governato, Tozzi \&
Cavalieri (1995) (GTC95) note that the effects of continuing infall of
surrounding galaxies onto an overdense region have been ignored in
earlier studies, and show that this can regenerate real compact
configurations after the galaxies in the initial collapsed group have
merged.

Given the difficulties of small number statistics and projection effects
associated with the galaxies in groups, an attractive option is to study
hot gas trapped in the group potential, rather than just the galaxies
themselves. In the case of galaxy clusters, this has proved a very
successful approach, and the gas has turned out to constitute the bulk of
the baryonic mass. The detection of significant quantities of hot gas
associated with a group as a whole, would provide strong evidence for a
genuine bound system, and one might hope to learn more about the dynamics
and evolutionary history of groups by studying the properties of the gas
in detail.

Given the velocity dispersion of compact groups ($\sim$100-400\kmps),
gas temperatures $\sim$0.1-1\keV\ are to be expected. Hence the gas
is best studied in the soft X-ray region of the spectrum.
Prior to the launch of \ROS\ such studies were not feasible. However,
the combination of high sensitivity, low background, and reasonable spatial
resolution provided by the \ROS\ Position Sensitive Proportional Counter
(PSPC) has changed this situation, and several studies of the X-ray properties
of individual groups (Mulchaey et al 1993; Ponman \& Bertram 1993; David \etal\
1994; Sulentic, Pietsch \& Arp 1995) and collections of groups (Ebeling, Voges
\& B\"ohringer 1994 (EVB94); Pildis, Bregman \& Evrard 1995 (PBE95);
Saracco \& Ciliegi 1995 (SC95); David, Jones \& Forman 1995; Mulchaey \etal\
1995; Henry \etal\ 1995) have now been published.

It is clear from these studies that some galaxy groups (both loose and compact)
do contain hot X-ray emitting gas with $T\sim 1$\keV, but beyond this, much of
the evidence about the properties of this emission, and its correlation with
other group properties, tends to be rather anecdotal, due to the lack of any
sensitive survey with a reasonable level of completeness. This paper
presents the results of the first such survey.

We have combined data from the \ROS\ All Sky Survey (RASS) with deeper
pointings, to produce an almost complete survey of the properties of the
Hickson compact galaxy groups. This allows us to study the X-ray luminosity
function of these groups, and to search for relationships between the X-ray
luminosity and temperature and other group properties. The data and their
reduction are described in section~2, and the results of the survey presented
in sections~3-5, after which, in section~6, we consider the implications of our
results for models of compact groups. Distances throughout the paper
are derived assuming $H_0=50$\kmpspMpc.

\section{DATA REDUCTION}
The original catalogue of 100 Hickson groups (Hickson 1982)
was reduced by Hickson \etal\ (1992) to 92 systems, by discarding those
found to contain fewer than three galaxies with accordant redshifts.
In addition, we have excluded from our analysis HCG60, HCG65 and HCG94,
which appear to correspond to the cores of galaxy clusters (EBV94;
Ramella \etal 1994; Ebeling, Mendes de Oliveira
\& White 1995), and HCG4 and HCG91, which are completely  dominated in
the {\it PSPC} by strong X-ray flux from active galaxies within the group.
HCG25 and HCG54 have also been excluded since they appear
to be contaminated by nearby point sources or background features
in the RASS data.

It has become apparent (\cf\ Ramella \etal\ 1994; Rood \& Struble 1994; EBV94)
that many compact groups are associated with larger scale
structures. We have therefore retained in our study groups such
as HCG5 and HCG48, which may be falling into clusters, and HCG58
which is apparently part of the larger galaxy group MKW10 (Williams 1985).

Our sample therefore consists of 85 groups, for all of which either
pointed or survey observations with the \ROS\ ~PSPC are available. This
corresponds to 92\% of the accordant Hickson groups. Some basic
properties of these systems and of our X-ray data for them are given in
Table~1. The first four columns, taken from Hickson \etal\ (1992) and
Hickson (1993), show the HCG number, redshift, total number of accordant
galaxies, and spiral fraction (treating S0 galaxies as non-spirals). In
the next column we give the line-of-sight velocity dispersion derived
from the values presented by Hickson \etal\ (1992), which were corrected
for measurement errors. We have adjusted this velocity dispersion by a factor
$\sqrt{N/(N-1)}$ (where $N$ is the number of group galaxies) to give an 
unbiased estimate.

All the accordant Hickson groups were observed by the \ROS\
~PSPC during the \ROS\ ~All-Sky Survey, but exposure times were
generally only a few hundred seconds. We have supplemented these
observations with a total of 32 pointed observations, which have
exposures typically an order of magnitude larger, and permit a much more
sensitive analysis. The remaining columns in Table~1 give the origin of
the data (survey or pointed), the exposure time,
and the radius within which X-ray flux was gathered (see below).

The use of pointed data does not in practice introduce a serious
selection bias into our results, since these observations were
selected by a number of proposers using a variety of different
criteria. Hence the pointed targets are {\it not},
more X-ray luminous than the average, for example. In nine cases,
the pointing is serendipitous (i.e. the HCG was not the intended
target), and in only three instances were groups known to be X-ray bright
in advance of a pointed observation being proposed. In these
cases we corrected for the bias when estimating 
the X-ray luminosity function (see the discussion in section~\ref{sec:xlf}).

\subsection{Pointed data}
\label{sec:point}
Pointed observations were reduced in a standard way. An initial PSPC
image was extracted, omitting data recorded during periods of high
background flux (typically a few percent of each observation) and point
sources identified using a likelihood source searching program (Allan,
Ponman \& Jeffries, in preparation). A spectral image (i.e. an $x,y,E$
data cube) covering the 0.1-2.4 keV band was then extracted for the
vicinity of the group, and an estimate of the background subtracted from
it. The latter was generated from a source-free region of the field (for
most groups this was taken from the annulus $r=0.6-0.7$\deg, excluding
any detected sources) and corrected to the position of the group using
the known vignetting function of the instrument. Finally the
background-subtracted spectral image was corrected for vignetting and
divided by the exposure time, giving a map of the spectral flux.

Unrelated point sources identified in the first stage of the procedure
were removed from the image to the 95\% radius of the point spread
function. In many groups, emission from individual
galaxies was apparent. Our aim in the present survey is to study the
X-ray emission from the hot intragroup gas -- we therefore excluded flux
from individual galaxies, except in the case of dominant ellipticals
which appeared to be at the centre of the X-ray emission, since in this
case it is more appropriate to identify the `galaxy' emission with the
hot gas in the core of the group potential (as seen in cD clusters). This
applies in the case of the dominant galaxies HCG12a, HCG37a, HCG42a,
HCG62a and HCG97a. In all other cases, galaxy emission was removed by
excising a circle of radius typically 1.2\arcm. Radial profiles of the
flux from individual galaxies were examined to establish that this radius
was satisfactory.

Except in the case of the two brightest groups, HCG62 and HCG97, it
was found that X-ray emission was undetectable outside a radius
corresponding to 200\kpc. This was therefore adopted as the standard
metric radius within which the signal and spectrum was evaluated
for each group. In the case of HCG62 and HCG97, a larger radius
of 500\kpc\ was used.

The detected count rate within the metric radius was extracted and
compared with the Poisson error from the background. Only sources
exceeding a 3 sigma detection threshold are regarded as having detectable
diffuse emission in the following analysis. For these groups, a spectrum
was extracted by collapsing the region of the spectral image cube (excluding
point sources) within the metric radius. This spectrum was then fitted with
a hot plasma model (Raymond \& Smith 1977) with an absorbing column
frozen at the value derived from radio surveys of galactic \hi\ (Stark
\etal 1992). The metallicity was generally left as a free parameter in
these fits, but was often poorly constrained. In some cases
it was necessary to fix its value (0.3 solar was adopted) in order to
get a stable fit.

EVB94 noted the presence of some wisps of extremely soft
emission in the vicinity of several Hickson groups, which they
tentatively identified with portions of old supernova remnants within our
own galaxy. We also encountered several cases in pointed data (HCG10,
HCG18 and HCG31) in which we recorded an extremely soft signal, confined
entirely to our bottom energy channel of 0.1-0.2\keV. In none of these
cases was the source bright enough to image in order to determine its
morphology. Due to the likelihood of contamination, we therefore treat
these sources as non-detections. In the case of HCG33, we see a clear
signal from the group, but this is accompanied by strong flux in the
lowest energy band which is inconsistent with the high column
(\nH$=1.85\times10^{21}$\pcm2) of the source. In this case we have
excluded the bottom energy channel from the spectral fitting.

The fitted spectral model has been used in each case, to derive a
bolometric X-ray luminosity for detected sources. The latter were
corrected for the portions masked out to remove contaminating point
sources, by replacing the `holes' in the data with interpolated flux
values. In the case of undetected sources, we have used a standard
spectral model with temperature $T=1$\keV, metallicity $Z=0.3$ solar, and
the appropriate galactic absorption column, to convert the flux limits to
bolometric luminosities. The resulting parameters are listed for each
group in Table~2.

\subsection{Survey data}
The RASS data at the positions of all accordant Hickson groups were
analysed by EVB94 using a technique, known as VTP
(Ebeling \& Wiedenmann 1993), specially adapted to searching for extended
emission. VTP gives an estimate of the source flux which is
independent of the projected shape of the emission region. The percentage
of the overall flux detected directly depends, however, on the surface
brightness of the source, and corrections assuming a model profile have
to be applied to recover the total flux. For the present study we have
avoided this model dependence (except for the case of HCG51 -- see below)
by proceeding in the same way as for the above analysis of the pointed
data -- extracting flux estimates in the 0.1-2.35\keV\ band from a region
of 200\kpc\ radius around the optical centre of each group.

The background in the vicinity of each group was derived from the VTP
analysis (which effectively fits a Poisson distribution to the observed
distribution of counts in the field, after removal of detected sources), and
this then defines the background noise in the exposure. A 3 sigma
detection threshold was applied, as in the pointed analysis, and a total
of 18 accordant groups detected. However, several of these are
contaminated by AGN or by probable foreground soft emission (see EVB94)
or point sources, and in 6 cases we have higher quality
pointed data. There remain five sources (HCG51,82,83,85 and 86) which are
detected in the RASS data, but for which no PSPC pointed data are
available. In addition, we derived 3 sigma count rate upper limits within
$r=200$\kpc\ for a further 48 groups.

The poor statistical quality of the RASS data does not permit
any detailed spectral analysis, so, for both detected sources and for
upper limits, we have converted count rates into bolometric luminosities
using the same spectral model as for the pointed upper limits: a
$T=1$\keV, $Z=0.3$ Raymond \& Smith plasma model, with the appropriate
Stark column (given in Table~2). One exception was made to the use
of a metric radius of 200\kpc\ in the RASS data; HCG51 is a very X-ray
luminous group, and flux was detected by VTP out to a radius of at least
300\kpc. In this case we have therefore used the count rate derived
by Ebeling \etal, which was corrected for flux missed by the VTP
algorithm using an assumed King profile. This is approximately double
the luminosity within $r=200$\kpc.

\section{SURVEY RESULTS AND PREVIOUS WORK}
The derived bolometric luminosities and spectral properties of the 85
groups surveyed are given in Table~2. Note that the luminosities
tabulated relate to {\it diffuse} emission from the groups only. Wherever
emission appears to be related to a particular group galaxy, it has been
removed (except in the case of dominant ellipticals) as discussed in
section~\ref{sec:point}. There are some groups for which the statistical
quality of the data, or the small size of the group, did not permit
emission from galaxies to be identified and removed. These systems (8 in
total) are flagged with a quality value $q=2$ in Table~2. Fully resolved
systems are given quality flag 1, and upper limits are flagged with a
`U'. In practice, the X-ray luminosities of most of the $q=2$ systems are
sufficiently high that the contamination from galaxy emission (typically
a few $\times 10^{40}$\ergps, judging from the resolved systems) is
likely to be small.

It is interesting to compare the results in Table~2 with those of the
smaller surveys of Hickson groups carried out by PBE95 and SC95. Each of
these sets of authors analysed PSPC data for 12 HCGs (PBE95 also included
the NGC2300 group of Mulchaey \etal\ in their sample). Both found some
groups to contain hot gas with $T\approx 0.7-1.0$\keV. However there are
significant differences between their results and ours.

The most striking difference relates to low surface brightness emission.
One of the main conclusions of PBE95 was that diffuse X-ray emission is
only detected in systems with spiral fractions less than 50\%, whereas we
detect diffuse flux in several systems with high spiral content. This is
a very important point, since it has a strong bearing on the questions of
the reality of compact groups and the origin of the hot gas, as
discussed in section~\ref{sec:disc} below. The most extreme example is
HCG16, a system containing {\it only} spiral galaxies, which was studied
by SC95. They concluded that there was no diffuse emission, whilst we
detect emission with a low temperature ($T=0.30$\keV), and a bolometric
luminosity which is 32 times their quoted 0.5-2.3\keV\ upper limit!

There appear to be two reasons for this dramatic difference in results
from the same data. SC95 (and also PBE95) tested for the
presence of diffuse emission in groups by smoothing an image
and looking for signs of extension. This is not a very
sensitive test for low surface brightness emission.
In addition, SC95 limited their data to the band 0.5-2.3\keV. Since we
find that spiral-rich groups with diffuse emission tend to have low
temperatures, discarding the lower energy channels will have significantly
reduced their sensitivity to this emission.

To convince the reader that there is, in fact, extended X-ray emission in
HCG16, we show in Fig.\ref{fig:h16rad} the radial profile of the PSPC
surface brightness about the centre of the optical group, after emission
from the individual galaxies (and unrelated point sources) has been
removed. In the present case, the emission can actually be seen in a
heavily smoothed image, as is shown in Fig.\ref{fig:h16im}, in which
an adaptive smoothing algorithm (which smoothes low surface brightness
areas more heavily than high brightness regions) has been applied to a
0.1-2.4\keV\ image of the group. Finally, in Fig.\ref{fig:h16spec} we
show the PSPC spectrum of the diffuse emission (flux from the galaxies
is excluded) with the best fitting hot plasma model overlaid.

In the case of PBE95, the authors appear to have been rather unlucky in their
choice of groups, since we also find no detectable diffuse emission in
their spiral-rich groups, despite our more sensitive test. In the
luminous groups like HCG62, in which PBE95 detect extended emission, it
is noticeable that their luminosities are considerably lower than ours.
This is partly due to the fact that they exclude a region around the
galaxies, whilst in the case of dominant central ellipticals we include
this as part of the group emission, as described in
section~\ref{sec:point}, but in addition, as they note themselves, the
results are sensitive to the background level selected. In the case of
HCG62, at least,
they have adopted a background value which appears to be too high.
An independent analysis of this group by David, Jones \& Forman
(1995) agrees well with our earlier results (Ponman \& Bertram 1993).

\section{DISTRIBUTIONS AND CORRELATIONS}

\subsection{Luminosity and temperature}
The diffuse luminosity of our detected systems, and 3$\sigma$ upper limits for
the remainder of our sample, are plotted against their redshifts in
Fig.\ref{fig:lz}. Although we do not detect any of the 
high redshift HCGs, it is interesting to note that a group as luminous
as HCG62 would have been detected at any redshift out to $z\sim 0.1$.
This suggests that the high redshift HCGs in our sample (remembering that we
have excluded three which we believe to be cluster cores) are unlikely
to be misidentified clusters.

The X-ray luminosities of detected groups appear to be uncorrelated with
either the number of galaxies in the group (which ranges from 3 to 7
within our detected sample) or the total optical luminosity. The lack of
correlation with richness is not surprising if groups have been subject
to significant merging, which would reduce the population, whilst
probably increasing the amount of intragroup gas (due to galaxy winds
driven by starburst activity), but the lack of any relationship between
X-ray and optical luminosity, shown in Fig.\ref{fig:lxlb}, is interesting.
It strongly suggests that the hot gas is associated not with the
galaxies, but with the potential well as a whole. It also indicates
that the bulk of the gas did not originate from the galaxies.

In Fig.\ref{fig:lxlbhist}, we show the distribution of $L_X/L_B$ for the
detected systems ($L_B$ is taken from Hickson \etal\ 1992, and corrected to
$H_0=50$\kmpspMpc). This should be compared with the typical value
$L_X/L_B\approx 10^{-4}$ for spiral and low luminosity elliptical field
galaxies, whilst even for very bright ellipticals the ratio barely
reaches \tpow{-3} (Canizares, Fabbiano \& Trinchieri 1987). Noting that
most of the HCGs with the lowest values of $L_X/L_B$ are dominated by
spiral galaxies, it is apparent that the diffuse X-ray emission is too
bright to be attributed to the member galaxies, even in the $q=2$ cases
where we have not been able to remove the galaxy contribution (unless
previously unrecognised low luminosity active galactic nuclei are present).

Approximately 6\% of the galaxies in Hickson groups are starburst
galaxies (Zepf 1993), and it is well known that such galaxies can show
extended X-ray emission arising from galactic winds (e.g. Fabbiano 1988).
In the case of HCG16, three of the galaxies were detected by IRAS
(Hickson \etal\ 1989), and two of them (HCG16c and HCG16d, on the left in
Fig.\ref{fig:h16im}) have the warm far infrared (FIR) colours
characteristic of starburst galaxies. However, comparison with other
starburst and merging systems shows that the extended emission apparent
in Fig.\ref{fig:h16im} cannot be attributed simply to galactic winds. In
normal starbursts like M82 and NGC253, detectable X-ray emission from
winds extends only $\sim 10$\kpc\ from the galaxy, and even in the most
luminous merging galaxies like Arp~220, it is limited to a few galactic
radii; whilst in HCG16, its total extent is almost $\sim 400$\kpc. In
addition, there is a strong correlation between FIR and X-ray
luminosities, which extends from normal to merging galaxies (Ponman \&
Read 1995). This predicts a luminosity in HCG16 which agrees well with
the X-ray luminiosity detected from the galaxies themselves, but is only
one quarter of the observed luminosity of the whole system.

The distribution of temperature for the 16 groups for which we have been
able to extract spectra of sufficient quality to fit a hot plasma model,
is shown in Fig.\ref{fig:thist}. Apart from two exceptionally cool
groups (HCG15 and HCG16), the temperatures span only a factor
of two, with most lying between 0.6\keV\ and 1\keV. This small range
cannot be due to a temperature selection effect. For typical absorbing
columns, the sensitivity of the \ROS\ PSPC does not diminish greatly
until $T$ drops below 0.3\keV, and there is little change in count rate
for a given emission measure as $T$ rises from 1 to 2\keV. It seems that
systems which look like compact groups are physically limited to a narrow
range in $T$ (apart from the cluster cores which we have rejected).
Hotter systems would presumably be richer, and would fail Hickson's
isolation criterion. In order to understand why cooler systems are not
detected in any numbers, it is helpful to look at the relationship
between X-ray temperature and luminosity.

The $L_X:T$ relation for our spectral subsample of 16 systems is shown in
Fig.\ref{fig:lt}. The two variables are significantly correlated: Kendall's
rank correlation coefficient (which is unit normal distributed) has a value
of $K=2.6$ ($P=0.01$ of chance occurrence, from a two tailed test), and
omitting the two $q=2$ sources from the spectral sample, this rises
to $K=3.2$ ($P=0.001$). A regression line has been fitted to the HCG
points in Fig.\ref{fig:lt}. It is important to allow for the errors
in both luminosity and temperature when regressing, since neither are
negligible, so we have used the doubly weighted regression technique
discussed by Feigelson \& Babu (1992), and made available through the
{\it ODRPACK} package. This gives the relationship between bolometric X-ray
luminosity and temperature:

$$ {\rm log} L_X = (43.17\pm 0.26) + (8.2\pm 2.7)\, {\rm log} T .$$

This relationship, which is very steep, is marked with its 1$\sigma$ error
envelope, in Fig.\ref{fig:lt}.

In order to compare the trends in $L$, $T$ and later $\sigma$, with those
seen in clusters, we have defined a set of clusters with high quality
spectral data, which are also plotted on the figure. These ten clusters
are those common to the samples of Edge \& Stewart (1991), from where we
take the bolometric luminosity and galaxy velocity dispersion, and
Yamashita (1992) who gives temperatures and metallicities derived with
Ginga and (in the case of the the Perseus cluster) Tenma. The relevant
parameters for these clusters are shown in Table~3.

A best fit to the cluster $L:T$ relation has recently been derived
by White \etal (in preparation) by applying doubly weighted regression
to a large sample of clusters. They find, in our units,
$ {\rm log} L_X = 42.86 + 2.81\, {\rm log} T $, which is marked as a heavy dashed
line in Fig.\ref{fig:lt}. This line passes above most of the galaxy groups,
and is much flatter than the best fit
trend for the HCGs. Fitting a regression line to the sample
of clusters plus HCGs shown in Fig.\ref{fig:lt}, gives the relation:
$$ {\rm log} L_X = (42.62\pm 0.13) + (3.29\pm 0.17)\, {\rm log} T ,$$
which is marked as the heavy solid line in the figure.

The steep $L:T$ relation for groups suggests a reason for the
lack of many cool groups in our sample. There is clearly a great deal of
scatter about the mean trend, and the two low $T$ groups which are detected
both fall well above the regression line. {\it Typical} groups with
$T<0.5$\keV, would according to the regression, have luminosities
below our detection limit of $\sim$\tpow{41}\ergps.

\subsection{Galaxy morphology}
As has already been noted by the authors of previous studies
of the X-ray properties of samples of groups (EVB94; PBE95; SC95;
Mulchaey \etal\ 1995; Henry \etal\ 1995), there is a 
relationship between X-ray luminosity and galaxy morphology. In
Fig.\ref{fig:esl}, we show the distribution of luminosity for detected
groups, broken down into two classes according to whether the brightest
galaxy is of early or late type. It can be seen that the brightest
systems (those with $L_X>4\times$\tpow{42}\ergps) are {\it all} dominated
by early-type galaxies.

To assess the significance of any difference between the diffuse X-ray
luminosities of HCGs with brightest Sp and E/S0 galaxies, it is desirable
to use not only the luminosities of detected systems, but also the upper
limits. This can be done using `survival analysis', provided that the
sampling is `random', in the sense that the limits are unrelated to the
actual values of the parameter being censored. For example, this
requirement would {\it not} be met in a case where longer observations
are made for fainter sources, in an effort to detect them. Since we are
dealing with luminosities, the actual limits are determined by both
exposure time and source distance.  In the case of our survey, the
observation times were determined in almost all cases without any
knowledge of the X-ray brightness of particular sources, and in the
absence of strong evolution, the source distance is also unrelated to
intrinsic luminosity. Random censoring is therefore a good assumption.

We have used the {\it ASURV} package (Feigelson \& Nelson 1985; Isobe,
Feigelson \& Nelson 1986), which computes the significance of the
difference between two distributions using a rank statistic, which
reduces to the Wilcoxon statistic in the uncensored case. Three different
options are available for the weights attributed to the censored values:
the Gehan, logrank, and Peto \& Prentice statistics (Feigelson \& Nelson
1985). These tests are most reliable when the two samples being compared
are comparable in size, and are censored in similar ways, both of which
are true in the present case.  All three statistics find the luminosity
distributions of HCGs dominated by early and late type galaxies to
differ, with significances of 99\%, 96\% and 98\% respectively, for the
three tests.

As far as the morphological mix in the groups' galaxy population is
concerned, we find only a very weak anticorrelation ($K=-1.3, P=0.2$)
between $L_X$ and spiral fraction, \fsp, in our sample,
as can be seen in Fig.\ref{fig:ltfsp}.
This contrasts with earlier strong statements made on the basis of smaller
samples.
Fig.\ref{fig:ltfsp} shows
a considerably stronger relationship ($K=-2.3, P=0.02$) between
\fsp\ and $T$. Since $T$ indicates the depth of the potential, whilst
$L_X$ reflects the amount of gas trapped in it, these results suggest that
galaxy morphology is related to the overall mass density in a group.

\subsection{Velocity dispersion}
\label{sec:vdisp}
A fairly strong relationship between luminosity and velocity dispersion
($K=1.8, P=0.07$ -- rising to $K=2.5, P=0.01$ if the eight $q=2$ groups
are excluded) can be seen in Fig.\ref{fig:hcglv}. The plotted errors in
$\sigma$ are dominated by the statistical uncertainty in determining a
velocity dispersion from only a handful of galaxies.  Again, we can
obtain a better indication of the significance of this relationship by
taking into account the censored luminosity values shown in the figure.
Using both the Cox proportional hazard model, and the generalised Kendall
rank correlation statistic, adapted for censored data as described by
Isobe, Feigelson \& Nelson (1986), we find the correlation between $L_X$
and $\sigma$ to be significant at $>99.9\%$ confidence from either test.

A 2D regression on log$L_X$ and log$\sigma$, using the errors in both,
gives $$ {\rm log} L_X = (30.0\pm 5.1) + (4.9\pm 2.1)\, {\rm log} \sigma
,$$ using all 22 detected HCGs. As can be seen from Fig.\ref{fig:lv},
this is in reasonable agreement with an extension of the trend found from
clusters, shown as a dashed line: \ $ {\rm log} L_X = 25.45 + 6.92\, {\rm
log} \sigma $, (White \etal\ in preparation); though some flattening of
the cluster trend may be indicated. A best fit through our group $+$
cluster sample gives $$ {\rm log} L_X = (27.6\pm 1.4) + (5.9\pm 0.5)\,
{\rm log} \sigma .$$

This result may be compared to the report of Dell'Antonio, Geller \&
Fabricant (1994), from a study of X-ray and optical results for groups
and clusters, of marginal evidence that the $L:\sigma$ relation may
flatten below log$\sigma\approx 2.5$.  Dell'Antonio \etal\ note that they
did not remove the contribution of galaxies to the X-ray emission, and
suggest that this may be the cause of the change in slope.  We {\it have}
removed the galaxy contribution, and although our best fit to the group
trend is somewhat flatter than the cluster relation, the difference is
not statistically significant.

There is some indication ($K=1.6, P=0.1$) of a correlation between
$\sigma$ and $T$ in the HCG spectral sample (Fig.\ref{fig:vt}), and the
comparison with clusters makes the trend clear. The $\sigma:T$ relation
for clusters is consistent with a value for $\beta$, the ratio of
specific energy in galaxies to that in the gas, of unity. However, it is
clear that most groups have $\beta<1$, i.e. they fall to the high $T$/low
$\sigma$ side of the $\beta=1$ line. The median value of $\beta$
for our HCG sample is 0.47 -- the specific energy in the gas is higher
than that in the galaxies. The trend towards lower $\beta$ in smaller
systems is in agreement with the results of Bird, Mushotzky \& Metzler
(1995), who study a sample of clusters specially corrected for the effects
of substructure on velocity dispersion, and find $\sigma\propto
T^{0.61\pm0.13}$. A regression through our group $+$ cluster sample gives
$$ {\rm log} \sigma = (2.54\pm 0.04) + (0.55\pm 0.05)\, {\rm log} T ,$$
which is shown as a solid line in Fig.\ref{fig:vt}.

\subsection{Metallicity}
As can be seen in Table~2, the metallicity derived in cases where
statistics were sufficiently good to allow a fit, is in many cases
considerably lower than the value of 0.3-0.4 solar found in most rich
clusters. The distribution of derived HCG metallicities is shown in 
Fig.\ref{fig:zhist}, but note from Table~2 that the values above 0.4
mostly have large errors, and the overall weighted mean metallicity of the
sample is only 0.18 solar.

This result is surprising given the trend seen in clusters (e.g.
Arnaud 1994) for {\it higher} metallicity in cooler systems, which
would lead one to expect typical metallicities $\approx$0.6~solar in
systems with $T\sim 1$\keV. The low metallicities found here are in
accord with most previous ROSAT studies of hot gas in galaxy groups
(see e.g. the compilation in Mulchaey et al 1995). 
However, we believe that at present this result should be viewed with caution,
since ROSAT is unable to resolve individual emission lines, and inferred
metal abundances rely heavily on the assumption that the continuum is
correctly modelled as an isothermal hot plasma. When temperature
variations in the gas are taken into account, metallicities several times
higher are found to be consistent with the data (Bourner \& Ponman, in
preparation). On the other hand, in the poor group containing NGC2300,
the low abundance ($\approx$0.1~solar) can be resolved over a range of
radial elements, and is also supported by ASCA observations, which have
the advantage of much higher spectral resolution (Davis \etal\ 1995).

\section{THE X-RAY LUMINOSITY FUNCTION}
\label{sec:xlf}
The HCG catalogue starts to become incomplete for total group magnitudes
$B_T>14$ (Hickson, Kindl \& Auman 1989), and the high redshift tail of
the catalogue represents a biased subset with particularly luminous
galaxies, high velocity dispersion and low spiral fraction (Whitmore
1990). It is also more difficult to determine galaxy morphology in the
higher redshift groups, and all three of the groups we reject as being
likely cluster cores lie at $z>0.04$. To avoid possible biases, we
therefore base our analysis of the X-ray luminosity function of HCGs on
the subsample of 62 of the 85 groups in our survey with $z<0.04$. We
detect diffuse emission from 18 of these systems. Since we know the
distribution of 3$\sigma$ upper limits for this set of 62 pointings, we
can estimate the fraction of groups in each luminosity bin which show
detectable diffuse emission as $$ f(L_i) = N(L_i)/N_{lim}(\leq L_i) ,$$
where $N(L_i)$ is the number of detected systems in the $i$th luminosity
bin, and $N_{lim}(\leq L_i)$ is the number of groups in the sample with
3$\sigma$ detection limits $\leq L_i$. 

We applied a small
correction to $N_{lim}(\leq L_i)$, to allow for the fact that three of
the long observations, of HCG42, HCG92 and HCG97, were made with {\it
prior knowledge} that the groups were bright. These three observations
were accordingly not available for the detection of {\it faint} systems,
and to include them in $N_{lim}(\leq L_i)$ for low values of $L_i$
would therefore give an underestimate of $f(L_i)$. This has been allowed
for by including these three pointings in
$N_{lim}(\leq L_i)$ as though the limiting sensitivity of each was
the actual luminosity of the group concerned.

The resulting frequency distribution for a series of logarithmic
luminosity bins is shown in Fig.\ref{fig:lfreq}a, and the cumulative
distribution in Fig.\ref{fig:lfreq}b.
The lowest populated luminosity bin plotted is likely to underestimate the
true frequency of occurrence of diffuse emission, since one should
really divide $N(L_i)$ by the number of systems with detection thresholds
for {\it diffuse} emission $\leq L_i$, whereas we are using detection
thresholds for {\it total} X-ray emission. In the lowest bin, the galaxy
contribution to $L_X$ may be quite significant, so we overestimate
$N_{lim}(\leq L_i)$, and underestimate $f(L_i)$.

From Fig.\ref{fig:lfreq}b, our best estimate of the total fraction of
HCGs with diffuse luminosity $L_X>10^{41.1}$\ergps, is $0.86\pm0.25$,
where the error arises from Poisson fluctuations in the number of
detected systems, and is dominated by uncertainties in the frequency of
low $L_X$ systems. The fraction of HCGs with luminosities in excess of
\tpow{42}\ergps is $0.20\pm0.07$. 
Monte Carlo simulations, using sources with luminosities drawn at random
from a known distribution, confirmed that the procedure above gives an
unbiased estimate of the fraction of systems with luminosity above
a chosen threshold, and that the calculated errors are realistic.

Alternatively, if one makes the assumption of random
censoring (see section~4.2) then the luminosity function can be estimated 
using the Kaplan-Meier estimator (Feigelson \& Nelson 1985). This gives 
a value of 0.75 for the fraction of groups with $L_X>10^{41.1}$\ergps.

One respect in which the sample of accordant Hickson groups is less than
satisfactory, is that groups with 3 or more accordant galaxies have been
retained in the sample, even though Hickson's original selection
criterion required four members. (For consistency one should really go
back and repeat the whole search process with a threshold of three member
galaxies.) If we exclude those groups in our sample with only three
accordant members, we reduce the sample at $z<0.04$ from 62 to 45, of
which 17 are detected. Repeating the analysis as before, we estimate that
the fraction of HCGs groups with $\geq4$ accordant members which have
diffuse emission above our threshold of $L_X=10^{41.1}$\ergps\ is
$1.06\pm0.32$ -- i.e. all of them! An analysis using the Kaplan-Meier
estimator gives an estimated fraction of 0.85 above the threshold in this
case. These large fractions have important implications for the reality
of compact groups, as will be discussed in the next section.

Mendes de Oliveira \&Hickson (1991) have
used the statistically homogenous sample of HCGs with $\geq4$ accordant
members to estimate a space density for these systems. Using simulations
to correct for selection effects, they obtain a density of
$4.9\times$\tpow{-6}\Mpc$^{-3}$ (for $H_0=50$). We can use this,
together with our results, to construct an X-ray luminosity function
for these systems. This is shown in Fig.\ref{fig:xlf}, and compared to
two estimates of the cluster luminosity function, extrapolated down
to low luminosity systems.

The best available estimate of the local cluster X-ray luminosity
function (XLF), is that derived from the flux limited \ROS\ Brightest
Cluster Sample, by Ebeling \etal\ (1996). This is well described by a
Schechter function, which is shown as a solid line in Fig.\ref{fig:xlf}.
Also shown (dotted) is the 1$\sigma$ error envelope for the simple power
law cluster XLF derived from the smaller sample of clusters available
from the {\it Einstein} Medium Sensitivity Survey (the lowest redshift
shell, $z=0.14$-0.20) by Henry \etal\ (1992). Note that the Henry \etal\
function has to be extrapolated by over an order of magnitude
in $L_X$ (the lowest luminosity in their cluster sample is
$\sim 10^{44}$\ergps) and hence has a rather large associated
error, whilst the Ebeling \etal\ sample extends down to $L_X \sim
10^{43}$\ergps.

The cluster results
actually apply to the energy bands 0.3-3.5\keV\ (Henry \etal) and
0.1-2.4\keV\ (Ebeling \etal), whilst our luminosities are bolometric;
however, for typical HCG spectra, restriction to either of these bands
would reduce our luminosities by only $\sim20$-30\%. Henry \etal\ (1995)
have recently derived a point on the XLF for loose galaxy groups, using
RASS data from a region around the North Ecliptic Pole, where the survey
exposure was greatest. This point (plotted in Fig.\ref{fig:xlf}) fits
quite well on an extrapolation of the Henry \etal\ (1992) XLF, however it
based on a very small sample (3 groups) and its statistical error
gives an overoptimistic impression of its reliability, due to the
possible influence of large
scale structure. Larger surveys of loose groups are required to establish
the behaviour of the cluster XLF for poor systems.

We have determined the slope of the HCG luminosity function by regressing
on the errors in $n(L)$, omitting the lowest bin, which is expected to be
underestimated, as explained above. This gives $$ {\rm log} n(L) =
(69.0\pm 9.9) - (1.74\pm 0.23)\, {\rm log} L_X ,$$ with a slope very
similar to that ($\alpha=1.78\pm0.09$) of Ebeling \etal, but marginally
flatter than that ($\alpha=2.19\pm 0.21$) derived by Henry \etal\ (1992).
Comparing with the more reliable XLF of Ebeling \etal, it appears that
the HCGs account for approximately 4\% of the X-ray luminosity of galaxy
groups as a whole. For comparison, Mendes de Oliveira \&Hickson (1991)
estimate that galaxies in HCGs account for approximately 0.8\% of the
total galactic light in the local Universe.

\section{DISCUSSION}
\label{sec:disc}

HCGs are undoubtedly to some extent a heterogeneous collection. With the
availability of galaxy redshifts, eight of the original 100 groups have already
proved to be chance projections (Hickson \etal\ 1992), and several more appear
to be really cluster cores. The interesting question is, what are {\it most} of
them?  The observed peculiarities of HCG properties demonstrate that they
cannot be chance projections of field galaxies, and theoretical studies show
that rather few of them are likely to be isolated bound configurations with
ages of more than a few Gyr, since this should result in very substantial
merging, leading to a large dominant merger remnant (e.g. Barnes 1989) which is
not generally observed (Mamon 1995).

However, there remain at least four credible pictures against which we can
measure our results. Most HCGs could be real bound systems which are (i) created
by collapse of subsystems within larger groups (DGR94, DGR95), or (ii) are
single systems refreshed by continuing infall of new galaxies from the
surrounding field (GTC95). Alternatively, they could mostly be (iii) chance
projections of galaxies within looser groups or clusters (Mamon 1986), or
(iv) unbound filamentary structures viewed along their length (HKW95).

The most immediately striking result of the survey reported above is the
high fraction ($>75$\%) of HCGs inferred to show diffuse X-ray emission,
once one corrects for visibility factors in a largely complete sample.
This contrasts with the typical values of $\sim30$\% inferred in previous
studies from the detection rate without such correction.
This poses a problem for the `filament hypothesis'. If HCGs are mostly
filaments several Mpc in length, then the X-ray surface brightness we
observe, even in the {\it fainter} groups, requires the filaments to
contain hot gas with $T\sim 1$\keV\ and particle density
$n\sim10^{-4}$\pcc (the length of such a filament along our line-of-sight
is determined by the galaxy velocity dispersion, which in this picture
arises from the Hubble flow). HKW95 mention the possibility of shock
heated gas within the filaments, but give no details of the densities
expected.  Briel \& Henry (1995) have recently searched for X-ray
filaments between neighbouring clusters and obtain 2$\sigma$ upper
limits corresponding to $n=5\times10^{-5}-10^{-4}$\pcc, for assumed
temperatures of 0.5-1\keV.

The filament hypothesis also has no obvious explanation for the 
correlations between $L_X$, $T$, velocity dispersion and spiral fraction,
reported above. It appears, therefore, that it can be ruled out, except
for the $<25$\% of HCGs which show no diffuse X-ray emission.
Ostriker, Lubin \& Hernquist (1995) conclude from the relative luminosities
of groups and clusters in the X-ray and optical bands, that there is
evidence that either many apparently compact groups are elongated along
the line-of-sight, or that smaller systems are relatively gas poor.
However, David, Jones \& Forman (1995) show that the gas fraction
does indeed increase with system size, and in view of our results, this seems
the most likely explanation for the trends noted by Ostriker \etal.

The idea of chance alignments within loose groups is more difficult to
eliminate, since some loose groups also exhibit diffuse X-ray emission
with properties rather similar (Mulchaey \etal\ 1995)
to those of HCGs. The fact that
the HCG luminosity function is not very different in slope to the cluster-group
XLF is also consistent with the idea that one is seeing just a chance
subset of the loose groups, viewed at a fortuitous angle. Even if the
flatter slope of the HCG function were a robust result, one could always
argue that selection effects make the appearance of compact projections
more likely within loose groups with denser cores, which would probably
be more luminous than the average. However, under the projection hypothesis
the HCG galaxies are a random sample taken from the larger loose group
membership, and it therefore becomes difficult to explain why one sees
the rather pronounced relationships between X-ray and galaxy properties
manifested in the different
luminosity distributions for spiral and elliptical dominated HCGs
(Fig.\ref{fig:esl}), or the correlation between $T$ and \fsp\
(Fig.\ref{fig:ltfsp}b).

In addition, the lack of any relationship between
$L_X$ and $L_B$ suggests strongly that we must be seeing a genuine
potential well. This does not rule out a loose group, but it does
imply that it would have to be collapsed, whereas Mamon (1994) argues
that the great majority of loose groups are still in the early stages 
of collapsing.

Our results therefore lend strong support to the idea that most
HCGs are genuinely compact systems, though not necessarily fully virialised
(\cf\ the X-ray morphology of HCG16). In terms of discriminating between
the two models for the formation of such groups, we note that DGR95
predict from their models of compact group formation within looser
groups, that the $L:T$ and $L:\sigma$ relations for HCGs should be
much {\it flatter} than those seen in clusters -- they obtain 
$L\propto T^{0.68}\propto \sigma^{0.57}$. The relations we observe
are much steeper than this. However, DGR95's models lack
some important and relevant astrophysics, notably galaxy winds,
so it is not clear that this disagreement is an insuperable problem for
their basic picture, which agrees well with the observation that
many HCGs are embedded in larger structures.

The fact that HCGs seem to fall rather nicely onto the $L$:$\sigma$ trend
for clusters does suggest that they are most likely to be self-contained
virialised systems (albeit with continuing infall). However, there does
seem to be one major problem with the model of GTC95. It predicts the
presence of a dominant merger remnant, from the early collapse and merger
of the perturbation which seeded the present group. It has already been
pointed out by Mendes de Oliveira \& Hickson (1991) that the difference
in magnitude between first and second ranked galaxies in HCGs is no
different for groups with dominant ellipticals (such as would be produced
by merging) than for dominant spirals -- although it appears that
selection effects have played a part in reducing the number of systems
with dominant galaxies identified as HCGs (Prandoni, Iovino \&
MacGillivray 1994). In addition, we note that although the four brightest
HCGs in our detected sample have elliptical first ranked galaxies, in
only one of these (HCG51) is the brightest galaxy more than twice as
optically luminous as the second ranked galaxy. If the picture of a long
lasting virialised core is to be retained, then it appears that some way
round the merging instability needs to be found. On the other hand, the
fact that X-ray luminosity seems to be correlated rather strongly with the
morphology of the first ranked galaxy, but much more weakly to the
overall fraction of early-type galaxies in a group, does suggest that
{\it some} merging has occurred in the more luminous systems.

Finally, it is interesting to consider the possible reasons for the
steepening of the $L:T$ relation at $T\sim 1$\keV\ seen in
Fig.\ref{fig:lt}. Since low $T$ corresponds, through the virial theorem,
to a shallow potential well, the most obvious interpretation is that
energy injection from galaxies (which have internal velocity dispersions
similar to those of galaxies within a group) might have a substantial
effect on groups, whilst being insignificant in rich clusters. The main
effect of energy injection into the intracluster medium is to reduce the
gas density in the central regions, and hence the X-ray luminosity
(Metzler \& Evrard 1994). Since this effect is much more pronounced in
smaller mass systems, and some gas may even escape from the well
altogether (Renzini \etal\ 1993), the effect is to steepen the $L:T$
relation at low temperatures (Knight \& Ponman, in preparation). If this
interpretation is correct, then the large scatter about the mean $L:T$
trend apparent in Fig.\ref{fig:lt} must be due primarily to differences
in the wind injection history of different groups.

Vigorous wind injection also has a more modest effect on the gas temperature,
which is raised by the injection of extra high entropy gas. Bird,
Mushotzky \& Metzler (1995) suggest that this, together with some
reduction in velocity dispersion due to the more effective action of
dynamical friction in low mass systems, is responsible for the lower
value of $\beta$ in smaller systems, as reported in
section \ref{sec:vdisp}.

\section{SUMMARY}
From our survey, covering 92\% of the accordant Hickson compact groups,
we deduce that at least three quarters of these systems exhibit detectable
X-ray emission from hot intragroup gas. In most cases, the
luminosity of this diffuse emission is considerably higher than that
arising from individual galaxies.
We find that the HCGs account for $\approx$4\% of the X-ray luminosity
density of galaxy groups, and have a
luminosity function similar to the cluster XLF.

We have investigated the spectral properties of the diffuse X-rays, and
find that the temperature is confined to a rather narrow band
(0.5-1.0\keV\ in most cases), even though the luminosity ranges over two
orders of magnitude. This represents a considerable steepening of the
cluster $L:T$ relation below $T\sim$1\keV, which may result from the
action of galaxy winds. 

The low metallicity apparent from our spectral fits (typically 0.2 solar
for our sample as a whole) is not yet a secure result, given the
possibility of a downward bias arising from fitting single temperature
models to multi-temperature gas. If it is confirmed by observations with
higher resolution X-ray spectrometers, then it implies that the majority
of the gas in the group wells is primordial, and the fact that it falls
below the metallicity of most rich clusters has interesting implications
for the evolution of galaxies and the intracluster medium, and will place
limits on the amount of wind injection which can be invoked.

The properties of the X-ray 
emission are not strongly related to the luminosity or number of galaxies
in a group, but they are related to galaxy morphology, with elliptical-rich
systems being typically both brighter and hotter. However, earlier reports
that spiral-rich systems show no diffuse X-ray emission are found to be
incorrect. Emission is certainly present in some such systems, but it
tends to be of lower surface brightness.

We argue that the high incidence of diffuse X-ray emission and the
strong correlations of $L_X$ with $\sigma$ and $T$ rule out the possibility
that most HCGs could be cosmic filaments viewed at a fortuitous
inclination, and strongly support the idea that they are genuinely
compact systems.

\section*{ACKNOWLEDGEMENTS}
We would like to thank Steve Zepf for his contributions to the early
stages of this project, David White for communicating results prior to
publication, and Gary Mamon and Pat Henry for helpful discussions.
This work made use of the Starlink facilities at Birmingham
and the LEDAS database at Leicester. PDJB acknowledges the receipt
of a PPARC studentship.

 \parskip=2.0truemm plus 0.5truemm       

\section*{REFERENCES}
Arnaud M., 1994, in Seitter W. ed., Cosmological Aspects of X-ray Clusters
of Galaxies. Kluwer, Dordrecht, p.197

Barnes J., 1984, MNRAS, 208, 873

Barnes J., 1985, MNRAS, 215, 517

Barnes J.E., 1989, Nat, 338, 123

Bird C.M., Mushotzky R.F., Metzler C.A., 1995, ApJ, in press

Briel U.G., Henry J.P., 1995, A\&A, 302, L9

Canizares C.R., Fabbiano G., Trinchieri G., 1987, ApJ, 312, 503

David L.P., Jones C., Forman W., 1995, ApJ, 445, 578

David L.P., Jones C., Forman W., Daines S., 1994, ApJ, 428, 544

Davis D.S., Mulchaey J.S., Mushotzky R.F., Burstein D., 1995, ApJ, in press

Dell'Antonio I.P., Geller M.J., Fabricant D.G., 1994, AJ, 107, 427

Diaferio A., Geller M.J., Ramella M., 1994, AJ, 107, 868 (DGR94)

Diaferio A., Geller M.J., Ramella M., 1995, AJ, 109, 2293 (DGR95)

Ebeling H., Allen S.W., Crawford C.S., Edge A.C., Fabian A.C., 
B\"ohringer H., Voges W., Huchra J.P., 1996, in proceedings of the
W\"urzburg conference:{\it R\"ontgenstrahlung from the Universe}, in press

Ebeling H., Mendes de Oliveira C., White D.A., 1995, MNRAS, in press

Ebeling H., Voges W., B\"ohringer H., 1994, ApJ, 436, 44 (EVB94)

Ebeling H., Wiedenmann G., 1993, Phys.Rev.E, 47, 704

Edge A.C., Stewart G.C., 1991, MNRAS, 252, 428

Fabbiano G., 1988, ApJ, 330, 672

Feigelson E.D., Babu G.J., 1992, ApJ, 397, 55

Feigelson E.D., Nelson P.I., 1985, ApJ, 293, 192

Governato F., Tozzi P., Cavalieri A., preprint (GTC95)

Henry J.P., Gioia I.M., Maccacaro T., Morris S.L., Stocke J.T, Wolter A.,
1992, ApJ, 386, 408

Henry J.P. \etal, 1995, ApJ, 449, 422

Hernquist L., Katz N., Weinberg D.H., 1995, ApJ, 442, 57 (HKW95)

Hickson P., 1982, ApJ, 255, 382

Hickson P., 1993, Ap.Lett.\& Comm., 29, 1

Hickson P., Kindl E., Huchra J.P., 1988, ApJ, 331, 64

Hickson P., Kindl E., Auman J.R., 1989, ApJS, 70, 687

Hickson P., Mendes de Oliveira C., Huchra J.P., Palumbo G.G.C.,
1992, ApJ, 399, 353

Hickson P., Menon T.K., Palumbo G.G.C., Persic M., 1989,
ApJ, 341, 679

Isobe T., Feigelson E.D., Nelson P.I., 1986, ApJ, 306, 490

Mamon G.A., 1986, ApJ, 307, 426

Mamon G.A., 1987, ApJ, 321, 622

Mamon G.A., 1992a, in Thuan T.X., Balkowski C., Tran Thanh Van J. eds.,
Physics of Nearby Galaxies: Nature or Nurture? Editions Frontieres, 
Gif-sur-Yvette, p.367

Mamon G.A., 1992b, ApJ, 401, L3

Mamon G.A., 1994, in Durret F., Mazure A., Tran Thanh Van J. eds.,
Clusters of Galaxies. Editions Frontieres, Gif-sur-Yvette, p.291

Mamon G.A., 1995, in Richter O.G. \& Borne K., eds., Groups of Galaxies.
ASP, San Francisco, p.83

Mendes de Oliveira C., Hickson P., 1991, ApJ, 380, 30

Mendes de Oliveira C., Hickson P., 1994, ApJ, 427, 684

Metzler C.A., Evrard A.E., 1994, ApJ, 437, 564

Mulchaey J.S., Davis D.S., Mushotzky R.F., Burstein D., 1993, ApJ, 404, L9

Mulchaey J.S., Davis D.S., Mushotzky R.F., Burstein D., 1995, ApJ, in press

Ostriker J.P., Lubin L.M., Hernquist L., 1995, ApJ, 444, L61

Pildis R.A., Bregman J.N., Evrard A.E., 1995, ApJ, 443, 514 (PBE95)

Ponman T.J., Bertram D., 1993, Nat, 363, 51

Ponman T.J., Read A.M., 1995, in Kunth D., Guideroni B., Heydari-Malayeri
M., Thuan T.X., Tran Thanh Van J. eds., The Interplay Between Massive
Star Formation, the ISM and Galaxy Evolution. Editions Frontieres, in
press.

Prandoni I., Iovino A., MacGillivray H.T., 1994, AJ, 107, 1235

Ramella M., Diaferio A., Geller M.J., Huchra J.P., 1994, AJ, 107, 1623

Raymond J.C., Smith B.W., 1977, ApJS, 35, 419

Renzini A., Ciotti L, D'Ercole A., Pellegrini S., 1993, ApJ, 419, 52

Rubin V.C., Hunter D.A., Ford W.K., 1991, ApJS, 76, 153

Rood H.J., Struble M.F., 1994, PASP, 106, 413

Saracco P., Ciliegi P., 1995, A\&A, 301, 348 (SC95)

Stark A.A., Gammie C.F., Wilson R.W., Bally J., Linke R.A., Heiles C.,
Hurwitz M., 1992, ApJS, 79, 77

Sulentic J.W., Pietsch W., Arp H., 1995, A\&A, 298, 420

Tully R.B., 1987, ApJ, 321, 280

Williams B.A., 1985, ApJ, 290, 462

Williams B.A., Rood H.J., 1987, ApJS, 63, 265


White S.D.M., 1990, in Wielen R ed., The Dynamics and Interactions of
Galaxies. Springer, Berlin, p.380

Whitmore B.C., 1990, in Oegerle W.R., Fitchett M.J., Danly L. eds.,
Clusters of Galaxies. CUP, Cambridge, p.139

Yamashita K., 1992, in Tanaka Y., Koyama K. eds., Frontiers of X-ray
Astronomy. Universal Academy Press, Tokyo. p.475

Zepf S.E., 1993, ApJ, 407, 448

Zepf S.E., Whitmore B.C., 1993, ApJ, 418, 72

\newpage
\section*{TABLES}
\vspace{10mm}
{\bf Table 1.} Basic properties of the 85 groups surveyed and the data
obtained.\\[5mm]
\begin{minipage}{220mm}
\begin{tabular}{rcccccrc}
\hline
 HCG &   $z$  &   $N_{\rm gal}$ &  $f_{\rm sp}$ &  log$\sigma$ (km s$^{-1}$) &
Pointed/Survey & $t_{\rm exp}$ (s) &   $r$ (kpc) \\
\hline
   1 & 0.0339 &  4 & 0.500 & 1.94 &  S  &   396  &  200\\
   2 & 0.0144 &  3 & 1.000 & 1.73 &  P  & 18429  &  200\\
   3 & 0.0255 &  3 & 0.333 & 2.48 &  P  &  7856  &  200\\
   5 & 0.0410 &  3 & 0.333 & 2.24 &  S  &   283  &  200\\
   6 & 0.0379 &  4 & 0.500 & 2.45 &  S  &   387  &  200\\
   7 & 0.0141 &  4 & 0.750 & 1.98 &  S  &   300  &  200\\
   8 & 0.0545 &  4 & 0.000 & 2.71 &  S  &   450  &  200\\
  10 & 0.0161 &  4 & 0.750 & 2.38 &  P  & 13861  &  200\\
  12 & 0.0485 &  5 & 0.200 & 2.43 &  P  &  8784  &  200\\
  13 & 0.0411 &  5 & 0.200 & 2.25 &  S  &   447  &  200\\
  14 & 0.0183 &  3 & 0.667 & 2.61 &  S  &   437  &  200\\
  15 & 0.0228 &  6 & 0.500 & 2.66 &  P  & 14029  &  200\\
  16 & 0.0132 &  4 & 1.000 & 2.13 &  P  & 14852  &  200\\
  17 & 0.0603 &  5 & 0.000 & 1.99 &  S  &   262  &  200\\
  20 & 0.0484 &  5 & 0.000 & 2.48 &  S  &   450  &  200\\
  21 & 0.0251 &  3 & 0.667 & 2.12 &  S  &   259  &  200\\
  22 & 0.0090 &  3 & 0.667 & 1.10 &  P  & 12211  &  200\\
  23 & 0.0161 &  4 & 0.750 & 2.26 &  P  &  4119  &  200\\
  24 &  0.0305 &   5 &  0.200 &  2.32 &   S &    225 &    200\\
  26 &  0.0316 &   7 &  0.571 &  2.31 &   P &   4914 &    200\\
  27 &  0.0874 &   4 &  0.250 &  1.92 &   S &    350 &    200\\
  28 &  0.0380 &   3 &  0.333 &  1.78 &   S &    266 &    200\\
  29 &  0.1047 &   3 &  0.000 &  2.70 &   S &    246 &    200\\
  30 &  0.0154 &   4 &  0.750 &  1.86 &   S &    525 &    200\\
  31 &  0.0137 &   3 &  1.000 &  1.67 &   P &   2428 &    200\\
  32 &  0.0408 &   4 &  0.000 &  2.37 &   S &    391 &    200\\
  33 &  0.0260 &   4 &  0.250 &  2.24 &   P &   4012 &    200\\
  34 &  0.0307 &   4 &  0.500 &  2.56 &   S &    310 &    200\\
  35 &  0.0542 &   6 &  0.167 &  2.54 &   P &  15669 &    200\\
  37 &  0.0223 &   5 &  0.400 &  2.65 &   P &   4803 &    200\\
  38 &  0.0292 &   3 &  1.000 &  0.96 &   P &   3754 &    200\\
  39 &  0.0701 &   4 &  0.500 &  2.34 &   S &    321 &    200\\
  40 &  0.0223 &   5 &  0.600 &  2.21 &   P &   3642 &    200\\
  42 &  0.0133 &   4 &  0.000 &  2.38 &   P &  12791 &    200\\
  43 &  0.0330 &   5 &  0.600 &  2.39 &   S &    373 &    200\\
  44 &  0.0046 &   4 &  0.750 &  2.16 &   P &   4672 &    200\\
  45 &  0.0732 &   3 &  0.667 &  2.34 &   S &    626 &    200\\
  46 &  0.0270 &   4 &  0.000 &  2.57 &   S &    432 &    200\\
  47 &  0.0317 &   4 &  0.750 &  1.45 &   S &    448 &    200\\
  48 &  0.0094 &   3 &  0.333 &  2.55 &   P &  12358 &    200\\
  49 &  0.0332 &   4 &  0.750 &  1.35 &   S &    743 &    200\\
  50 &  0.1392 &   5 &  0.000 &  2.72 &   P &   1783 &    200\\
\hline
\end{tabular}
\vspace{3mm}
\end{minipage}
\newpage
{\bf Table 1.} continued\\[5mm]
\begin{minipage}{220mm}
\begin{tabular}{rcccccrc}
\hline
 HCG &   $z$  &   $N_{\rm gal}$ &  $f_{\rm sp}$ &  log$\sigma$ (km s$^{-1}$) &
Pointed/Survey & $t_{\rm exp}$ (s) &   $r$ (kpc) \\
\hline
  51 &  0.0258 &   5 &  0.400 &  2.42 &   S &    290 &   $>$300\\
  52 &  0.0430 &   3 &  1.000 &  2.32 &   S &    435 &    200\\
  53 &  0.0206 &   3 &  0.667 &  1.85 &   S &    451 &    200\\
  55 &  0.0526 &   4 &  0.000 &  2.34 &   S &    801 &    200\\
  56 &  0.0270 &   5 &  0.200 &  2.26 &   S &    431 &    200\\
  57 &  0.0304 &   7 &  0.429 &  2.45 &   P &   3226 &    200\\
  58 &  0.0207 &   5 &  0.600 &  2.25 &   P &  11412 &    200\\
  59 &  0.0135 &   4 &  0.750 &  2.33 &   S &    443 &    200\\
  61 &  0.0130 &   3 &  0.333 &  2.01 &   S &    492 &    200\\
  62 &  0.0137 &   4 &  0.000 &  2.51 &   P &  18028 &    500\\
  63 &  0.0311 &   3 &  1.000 &  2.01 &   S &    308 &    200\\
  64 &  0.0360 &   3 &  0.667 &  2.40 &   S &    314 &    200\\
  66 &  0.0699 &   4 &  0.000 &  2.54 &   S &    567 &    200\\
  67 &  0.0245 &   4 &  0.500 &  2.38 &   P &  16193 &    200\\
  68 &  0.0080 &   5 &  0.200 &  2.23 &   P &  11865 &    200\\
  69 &  0.0294 &   4 &  0.500 &  2.40 &   S &    424 &    200\\
  70 &  0.0636 &   4 &  1.000 &  2.13 &   S &    497 &    200\\
  71 &  0.0301 &   3 &  1.000 &  2.70 &   S &    515 &    200\\
  72 &  0.0421 &   4 &  0.250 &  2.48 &   S &    414 &    200\\
  73 &  0.0449 &   3 &  0.667 &  1.98 &   P &   4557 &    200\\
  74 &  0.0399 &   5 &  0.000 &  2.54 &   S &    285 &    200\\
  75 &  0.0416 &   6 &  0.500 &  2.46 &   S &    243 &    200\\
  76 &  0.0340 &   7 &  0.286 &  2.41 &   S &    283 &    200\\
  79 &  0.0145 &   4 &  0.250 &  2.18 &   P &   4890 &    200\\
  80 &  0.0310 &   4 &  1.000 &  2.49 &   S &   1290 &    200\\
  81 &  0.0499 &   4 &  0.250 &  2.25 &   S &    553 &    200\\
  82 &  0.0362 &   4 &  0.500 &  2.85 &   S &    675 &    200\\
  83 &  0.0531 &   5 &  0.400 &  2.70 &   S &    491 &    200\\
  84 &  0.0556 &   5 &  0.200 &  2.33 &   P &  37490 &    200\\
  85 &  0.0393 &   4 &  0.000 &  2.62 &   S &   2370 &    200\\
  86 &  0.0199 &   4 &  0.000 &  2.48 &   S &    290 &    200\\
  87 &  0.0296 &   3 &  0.667 &  2.01 &   S &    444 &    200\\
  88 &  0.0201 &   4 &  1.000 &  1.25 &   S &    421 &    200\\
  89 &  0.0297 &   4 &  1.000 &  1.54 &   S &    448 &    200\\
  90 &  0.0088 &   4 &  0.500 &  2.04 &   P &  13518 &    200\\
  92 &  0.0215 &   4 &  0.750 &  2.65 &   P &  15737 &    200\\
  93 &  0.0168 &   4 &  0.500 &  2.37 &   P &  14784 &    200\\
  95 &  0.0396 &   4 &  0.750 &  2.54 &   S &    443 &    200\\
  96 &  0.0292 &   4 &  0.750 &  2.16 &   P &   2423 &    200\\
  97 &  0.0218 &   5 &  0.400 &  2.61 &   P &  12050 &    500\\
  98 &  0.0266 &   3 &  0.000 &  2.16 &   S &    402 &    200\\
  99 &  0.0290 &   5 &  0.200 &  2.46 &   S &    364 &    200\\
 100 &  0.0178 &   3 &  1.000 &  2.00 &   S &    442 &    200\\
\hline
\end{tabular}
\vspace{3mm}
\end{minipage}
\newpage
{\bf Table 2.} X-ray properties of the surveyed groups. Distances
assume $H_0=50$ km s$^{-1}$ Mpc$^{-3}$ and $N_H$ values are derived from
radio surveys. Temperatures
and metallicities were fixed at 1.0 keV and 0.3 solar (suffix F)
in cases where data were inadequate to allow a fit. Errors are
one sigma, and upper limits three sigma.
The values
marked with a dagger were too poorly constrained
to allow errors to be derived. The quality flag is discussed in the text.
\\[5mm]
\begin{minipage}{220mm}
\begin{tabular}{rrccccc}
\hline
HCG & $d$ (Mpc) & $N_H$ ($10^{20}$cm$^{-2}$) & $T$ (keV) & $Z$ (solar) &
 Qual & log $L_X$ (erg s$^{-1}$) \\  
\hline
  1 &  203  &   3.67  &            1.0F  &           0.3F  &  U &  $<$42.42 \\
  2 &  86   &   3.70  &            1.0F  &           0.3F  &  U &  $<$41.28 \\
  3 &  152  &   3.18  &            1.0F  &           0.3F  &  U &  $<$41.71 \\
  5 &  245  &   4.05  &            1.0F  &           0.3F  &  U &  $<$42.54 \\
  6 &  227  &   3.50  &            1.0F  &           0.3F  &  U &  $<$42.43 \\
  7 &  85   &   2.23  &            1.0F  &           0.3F  &  U &  $<$42.03 \\
  8 &  327  &   3.95  &            1.0F  &           0.3F  &  U &  $<$42.55 \\
 10 &  97   &   4.53  &            1.0F  &           0.3F  &  U &  $<$41.43 \\
 12 &  291  &   3.69  &   0.89$\pm0.12$  &           0.3F  &  1 &  42.31$\pm0.08$\\
 13 &  246  &   3.30  &            1.0F  &           0.3F  &  U &  $<$42.40 \\
 14 &  110  &   2.07  &            1.0F  &           0.3F  &  U &  $<$42.02 \\
 15 &  137  &   2.72  &   0.44$\pm0.08$  &           0.3F  &  1 &  41.80$\pm0.12$\\
 16 &  79   &   2.05  &   0.30$\pm0.05$  &  0.00$\pm0.17$  &  1 &  41.68$\pm0.06$\\
 17 &  362  &   7.57  &            1.0F  &           0.3F  &  U &  $<$42.78 \\
 20 &  290  &   9.64  &            1.0F  &           0.3F  &  U &  $<$42.63 \\
 21 &  151  &   2.60  &            1.0F  &           0.3F  &  U &  $<$42.33 \\
 22 &  54   &   3.80  &            1.0F  &           0.3F  &  U &  $<$41.15 \\
 23 &  97   &   5.91  &            1.0F  &           0.3F  &  U &  $<$41.72 \\
 24 &  183  &   5.30  &            1.0F  &           0.3F  &  U &  $<$42.52 \\
 26 &  189  &   4.73  &            1.0F  &           0.3F  &  U &  $<$41.95 \\
 27 &  524  &   4.10  &            1.0F  &           0.3F  &  U &  $<$43.03 \\
 28 &  228  &   5.55  &            1.0F  &           0.3F  &  U &  $<$42.70 \\
 29 &  628  &   2.50  &            1.0F  &           0.3F  &  U &  $<$43.02 \\
 30 &  92   &   4.65  &            1.0F  &           0.3F  &  U &  $<$42.07 \\
 31 &  82   &   4.71  &            1.0F  &           0.3F  &  U &  $<$41.93 \\
 32 &  245  &   5.87  &            1.0F  &           0.3F  &  U &  $<$42.51 \\
 33 &  156  &  18.50  &   0.61$\pm0.30$  &           0.3F  &  1 &  41.77$\pm0.11$\\
 34 &  184  &  16.42  &            1.0F  &           0.3F  &  U &  $<$42.55 \\
 35 &  325  &   2.54  &   0.91$\pm0.18$  &  0.16$\pm0.23$  &  2 &  42.35$\pm0.11$\\
 37 &  134  &   1.93  &   0.67$\pm0.11$  &  0.17$\pm0.15$  &  1 &  42.12$\pm0.06$\\
 38 &  175  &   3.02  &            1.0F  &           0.3F  &  U &  $<$42.01 \\
 39 &  420  &   3.23  &            1.0F  &           0.3F  &  U &  $<$42.80 \\
 40 &  134  &   3.15  &            1.0F  &           0.3F  &  U &  $<$41.73 \\
 42 &  80   &   3.94  &   0.82$\pm0.03$  &  0.50$\pm0.24$  &  1 &  42.16$\pm0.02$\\
 43 &  198  &   3.30  &            1.0F  &           0.3F  &  U &  $<$42.40 \\
 44 &  28   &   1.98  &            1.0F  &           0.3F  &  U &  $<$40.84 \\
 45 &  439  &   0.81  &            1.0F  &           0.3F  &  U &  $<$42.60 \\
 46 &  162  &   2.94  &            1.0F  &           0.3F  &  U &  $<$42.25 \\
 47 &  190  &   3.79  &            1.0F  &           0.3F  &  U &  $<$42.33 \\
 48 &  56   &   4.53  &   1.09$\pm0.21$  &  0.69$\pm0.10$  &  2 &  41.58$\pm0.14$\\
 49 &  199  &   1.64  &            1.0F  &           0.3F  &  U &  $<$42.26 \\
 50 &  835  &   0.78  &            1.0F  &           0.3F  &  U &  $<$42.40 \\
\hline
\end{tabular}
\end{minipage}
\newpage
{\bf Table 2.} continued\\[5mm]
\begin{minipage}{220mm}
\begin{tabular}{rrccccc}
\hline
HCG & $d$ (Mpc) & $N_H$ ($10^{20}$cm$^{-2}$) & $T$ (keV) & $Z$ (solar) &
 Qual & log $L_X$ (erg s$^{-1}$) \\  
\hline
 51 &  155  &   1.21  &            1.0F  &           0.3F  &  2 &  42.99$\pm0.11$\\
 52 &  258  &   1.58  &            1.0F  &           0.3F  &  U &  $<$42.50\\
 53 &  124  &   1.70  &            1.0F  &           0.3F  &  U &  $<$42.17\\
 55 &  315  &   1.51  &            1.0F  &           0.3F  &  U &  $<$42.45\\
 56 &  162  &   1.11  &            1.0F  &           0.3F  &  U &  $<$42.23\\
 57 &  182  &   1.70  &   0.82$\pm0.27$  &  0.48$\pm0.88$  &  1 &  41.98$\pm0.21$\\
 58 &  124  &   3.10  &   0.64$\pm0.19$  &   0.0$\pm0.16$  &  1 &  41.89$\pm0.11$\\
 59 &  81   &   2.94  &            1.0F  &           0.3F  &  U &  $<$42.00\\
 61 &  78   &   1.68  &            1.0F  &           0.3F  &  U &  $<$41.91\\
 62 &  82   &   3.00  &   0.96$\pm0.04$  &  0.15$\pm0.06$  &  1 &  43.04$\pm0.03$\\
 63 &  186  &   5.82  &            1.0F  &           0.3F  &  U &  $<$42.49\\
 64 &  216  &   2.41  &            1.0F  &           0.3F  &  U &  $<$42.48\\
 66 &  419  &   1.24  &            1.0F  &           0.3F  &  U &  $<$42.67\\
 67 &  145  &   2.16  &   0.82$\pm0.19$  &  0.49$\pm0.53$  &  1 &  41.69$\pm0.10$\\
 68 &  48   &   0.90  &   0.54$\pm0.15$  &  0.43$\pm0.98$  &  1 &  41.27$\pm0.26$\\
 69 &  176  &   1.44  &            1.0F  &           0.3F  &  U &  $<$42.33\\
 70 &  381  &   1.24  &            1.0F  &           0.3F  &  U &  $<$42.63\\
 71 &  180  &   1.53  &            1.0F  &           0.3F  &  U &  $<$42.28\\
 72 &  252  &   2.34  &            1.0F  &           0.3F  &  U &  $<$42.54\\
 73 &  269  &   3.06  &   0.59$^{\dag}$  &   0.0$^{\dag}$  &  2 &  42.43$\pm0.24$\\
 74 &  239  &   4.01  &            1.0F  &           0.3F  &  U &  $<$42.67\\
 75 &  249  &   4.15  &            1.0F  &           0.3F  &  U &  $<$42.72\\
 76 &  204  &   3.82  &            1.0F  &           0.3F  &  U &  $<$42.66\\
 79 &  87   &   3.51  &            1.0F  &           0.3F  &  U &  $<$41.71\\
 80 &  186  &   2.42  &            1.0F  &           0.3F  &  U &  $<$42.16\\
 81 &  299  &   4.41  &            1.0F  &           0.3F  &  U &  $<$42.60\\
 82 &  217  &   1.74  &            1.0F  &           0.3F  &  2 &  42.29$\pm0.14$\\
 83 &  318  &   5.86  &            1.0F  &           0.3F  &  2 &  42.81$\pm0.12$\\
 84 &  333  &   3.61  &            1.0F  &           0.3F  &  U &  $<$41.92\\
 85 &  236  &   6.58  &            1.0F  &           0.3F  &  2 &  42.27$\pm0.10$\\
 86 &  119  &   8.56  &            1.0F  &           0.3F  &  2 &  42.32$\pm0.14$\\
 87 &  177  &   5.12  &            1.0F  &           0.3F  &  U &  $<$42.36\\
 88 &  121  &   4.60  &            1.0F  &           0.3F  &  U &  $<$42.18\\
 89 &  178  &   4.64  &            1.0F  &           0.3F  &  U &  $<$42.30\\
 90 &  53   &   1.50  &   0.68$\pm0.12$  &  0.20$\pm0.16$  &  1 &  41.48$\pm0.09$\\
 92 &  129  &   6.70  &   0.75$\pm0.08$  &  0.23$\pm0.51$  &  1 &  42.16$\pm0.04$\\
 93 &  101  &   4.70  &            1.0F  &           0.3F  &  U &  $<$41.34\\
 95 &  237  &   4.38  &            1.0F  &           0.3F  &  U &  $<$42.43\\
 96 &  175  &   4.61  &            1.0F  &           0.3F  &  U &  $<$42.11\\
 97 &  131  &   3.29  &   0.87$\pm0.05$  &  0.12$\pm0.04$  &  1 &  42.78$\pm0.02$\\
 98 &  159  &   2.81  &            1.0F  &           0.3F  &  U &  $<$42.27\\
 99 &  174  &   4.87  &            1.0F  &           0.3F  &  U &  $<$42.34\\
100 &  107  &   4.30  &            1.0F  &           0.3F  &  U &  $<$41.99\\
\hline
\end{tabular}
\end{minipage}
\newpage
{\bf Table 3.} Cluster data used in Figs.8, 12 and 13, taken from Edge \& Stewart (1991)
and Yamashita (1992).\\[5mm]
\begin{tabular}{lcccc}
\hline
Cluster & $z$ & log$\sigma$ (km s$^{-1}$) & $T$ (keV) & log$L_X$ (erg s$^{-1}$) \\
\hline
 Perseus   &  0.0184 &  3.106  &   6.08$\pm0.20$ & 45.362$\pm0.019$ \\
 A478      &  0.0882 &  2.956  &   6.70$\pm0.19$ & 45.683$\pm0.015$ \\  
 A496      &  0.0320 &  2.848  &   3.91$\pm0.06$ & 44.826$\pm0.071$ \\  
 A1060     &  0.0114 &  2.784  &   2.55$\pm0.04$ & 43.820$\pm0.046$ \\
 A1367     &  0.0215 &  2.915  &   3.50$\pm0.18$ & 44.246$\pm0.047$ \\  
 Virgo     &  0.0038 &  2.758  &   2.34$\pm0.02$ & 43.568$\pm0.071$ \\  
 Centaurus &  0.0109 &  2.768  &   3.54$\pm0.13$ & 44.167$\pm0.038$ \\  
 Coma      &  0.0232 &  3.004  &   8.11$\pm0.07$ & 45.204$\pm0.027$ \\  
 A1795     &  0.0621 &  2.888  &   5.34$\pm0.11$ & 45.292$\pm0.027$ \\  
 A2142     &  0.0899 &  3.094  &   8.68$\pm0.20$ & 45.777$\pm0.023$ \\  
\hline
\end{tabular}

\newpage
\section*{FIGURE CAPTIONS}
\begin{figure}[ht]
\vspace{11mm}
\caption{Radial profile of diffuse X-ray emission from HCG16,
with flux from the member galaxies excised.}
\label{fig:h16rad}
\end{figure}
\begin{figure}[ht]
\vspace{11mm}
\caption{Contours of X-ray surface brightness (0.1-2.4\keV) after
smoothing with an adaptive filter, superimposed on an optical image of the
group. The dashed circle has a radius of 200\kpc.}
\label{fig:h16im}
\end{figure}
\begin{figure}[ht]
\vspace{11mm}
\caption{PSPC spectrum of the diffuse emission from HCG16, with best
fitting ($T=0.3\keV$, $Z=0$) hot plasma model overlaid.}
\label{fig:h16spec}
\end{figure}
\begin{figure}[ht]
\vspace{11mm}
\caption{Bolometric X-ray luminosities and 3$\sigma$ upper limits
plotted against redshift for the 85 survey groups.}
\label{fig:lz}
\end{figure}
\begin{figure}[ht]
\vspace{11mm}
\caption{X-ray and blue luminosities for the 22 detected HCGs. Groups
with dominant early type galaxies are plotted as ellipses, whilst
those with dominant spirals are shown as crosses. Systems in which
the galaxies are not resolved from the diffuse emission (i.e. with
quality flag $q=2$) are shown dotted. Vertical extent
represents 1$\sigma$ error in $L_X$, but horizontal extent is arbitrary.}
\label{fig:lxlb}
\end{figure}
\begin{figure}[ht]
\vspace{11mm}
\caption{Distribution of X-ray/optical luminosity ratio for the detected
systems.~~~~~~~~~~~~~~~~~~~}
\label{fig:lxlbhist}
\end{figure}
\pagebreak
\begin{figure}[ht]
\vspace{11mm}
\caption{Distribution of derived plasma temperature for HCGs with adequate
statistics to allow a fit.}
\label{fig:thist}
\end{figure}
\begin{figure}[ht]
\vspace{11mm}
\caption{$L_X:T$ relation for HCGs (crosses) and a sample of clusters
(diamonds) -- $q=1$ (i.e. fully resolved) systems are shown as barred crosses.
Vertical and horizontal extent represent 1$\sigma$ errors.
The solid and dotted lines overlaid on the groups indicate the best fit
regression line and its 1$\sigma$ envelope. The bold solid line is a fit
to the whole group $+$ cluster sample, whilst the dashed line labelled CL
shows the relation derived for a large cluster sample by White \etal\ (in
preparation).}
\label{fig:lt}
\end{figure}
\begin{figure}[ht]
\vspace{11mm}
\caption{Luminosity distribution for HCGs dominated by (a) early and
(b) late type galaxies. The distribution for the full sample is overlaid
as a dotted line.}
\label{fig:esl}
\end{figure}
\begin{figure}[ht]
\vspace{11mm}
\caption{Relationship between spiral fraction and (a) X-ray luminosity,
and (b) temperature. Barred crosses denote high quality ($q=1$) groups.
Values of the K statistic for the full sample and the $q=1$ systems only
(in brackets) are given.}
\label{fig:ltfsp}
\end{figure}
\begin{figure}[ht]
\vspace{11mm}
\caption{$L_X:\sigma$ plot, including 3$\sigma$ upper limits, for the
full HCG sample. High quality ($q=1$) groups are plotted as barred crosses. 
Errors bars correspond to 1$\sigma$.
Values of the K statistic for the full detected subsample and the 
$q=1$ systems only (in brackets) are given.}
\label{fig:hcglv}
\end{figure}
\begin{figure}[ht]
\vspace{11mm}
\caption{$L_X:\sigma$ relation for HCGs (crosses) and clusters
(diamonds). Vertical and horizontal extent represent 1$\sigma$ errors.
$q=1$ groups are shown as barred crosses.
The solid and dotted lines overlaid on the groups indicate the best fit
regression line and its 1$\sigma$ envelope. The bold solid line is a fit
to the whole group $+$ cluster sample, whilst the dashed line labelled CL
shows the relation derived for a large cluster sample by White \etal\ (in
preparation).}
\label{fig:lv}
\end{figure}
\pagebreak
\begin{figure}[ht]
\vspace{11mm}
\caption{$\sigma:T$ relation for HCGs (crosses) and clusters
(diamonds). Vertical and horizontal extent represent 1$\sigma$ errors.
$q=1$ groups are shown as barred crosses.
The solid and dotted lines overlaid on the groups indicate the best fit
regression line and its 1$\sigma$ envelope. The bold solid line is a fit
to the whole group $+$ cluster sample, and the dotted line is the locus
along which $\beta=1$, i.e. specific energy in gas and galaxies is equal.}
\label{fig:vt}
\end{figure}
\begin{figure}[ht]
\vspace{11mm}
\caption{Distribution of metallicity for HCGs in which it was possible to
fit for it.~~~~~~~~~~~~~~~~~~~~~}
\label{fig:zhist}
\end{figure}
\begin{figure}[ht]
\vspace{11mm}
\caption{(a) Estimated frequency distribution of bolometric X-ray luminosity 
for the sample of 62 systems with $z<0.04$ (see text for details), and 
(b) the corresponding cumulative distribution.}
\label{fig:lfreq}
\end{figure}
\begin{figure}[ht]
\vspace{11mm}
\caption{Normalised luminosity function for HCGs with $z<0.04$ and
$\geq4$ accordant galaxies. The dashed line is the best power law fit to
these points (excluding the lowest bin, which is probably
underestimated). Also shown are the cluster XLF of Ebeling \etal\ 1996
(solid line), and the 1$\sigma$ error envelope
for the power law XLF of Henry \etal\ 1992 (dotted), as well as the loose
group space density derived by Henry \etal\ 1995. Vertical error bars are
1$\sigma$.}
\label{fig:xlf}
\end{figure}
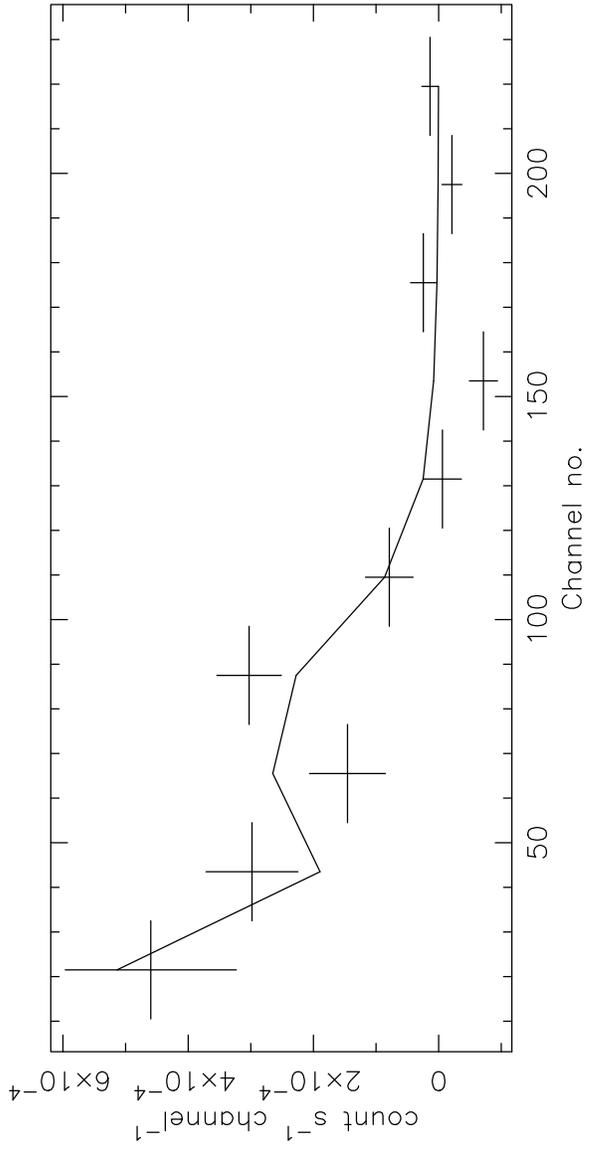
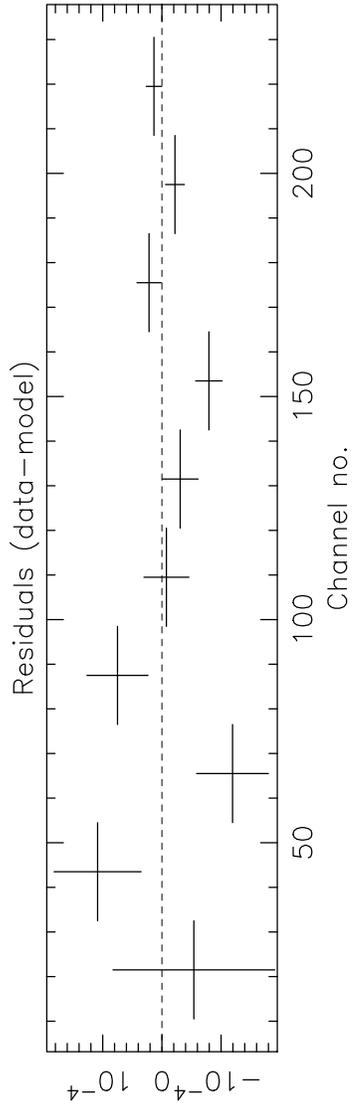
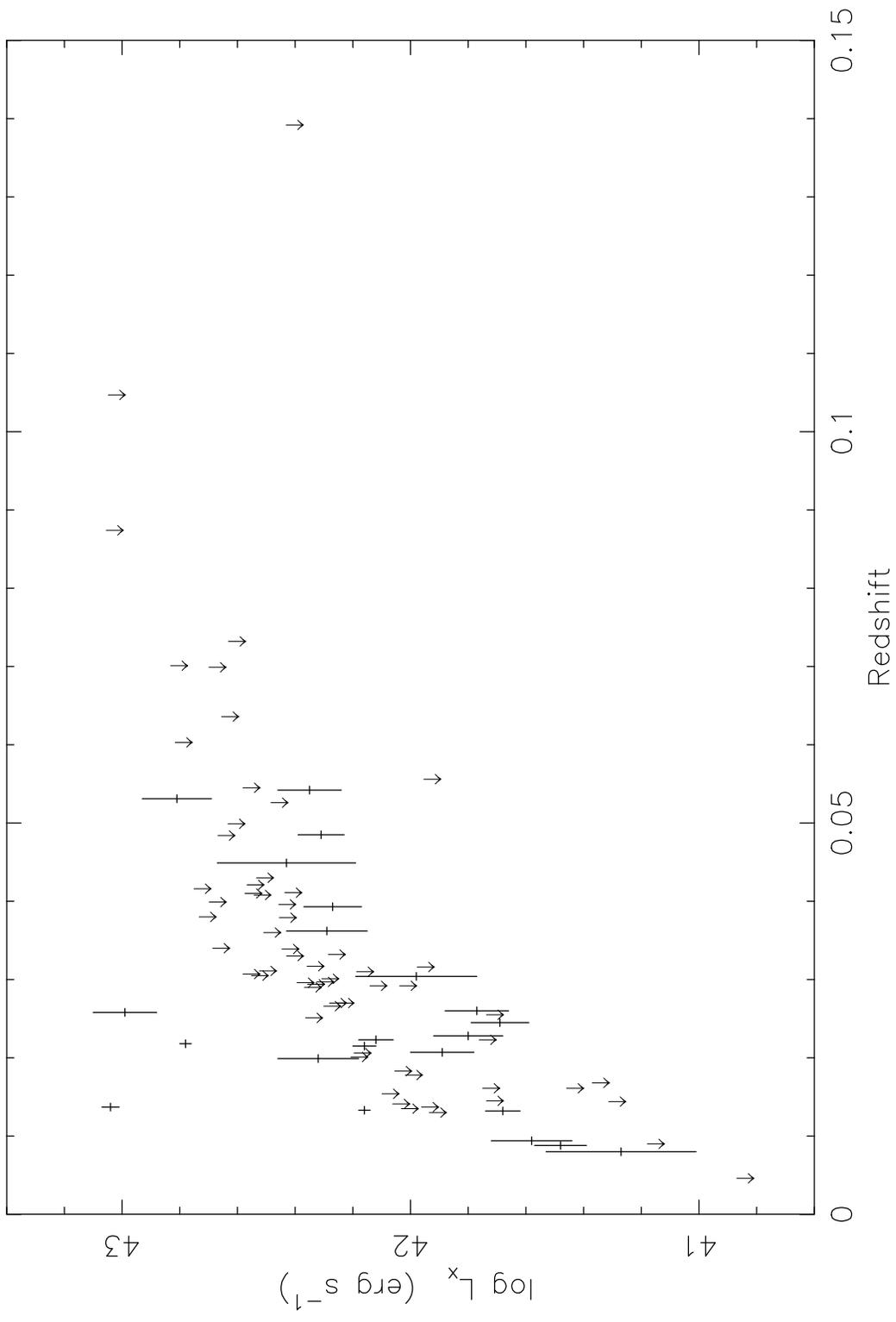
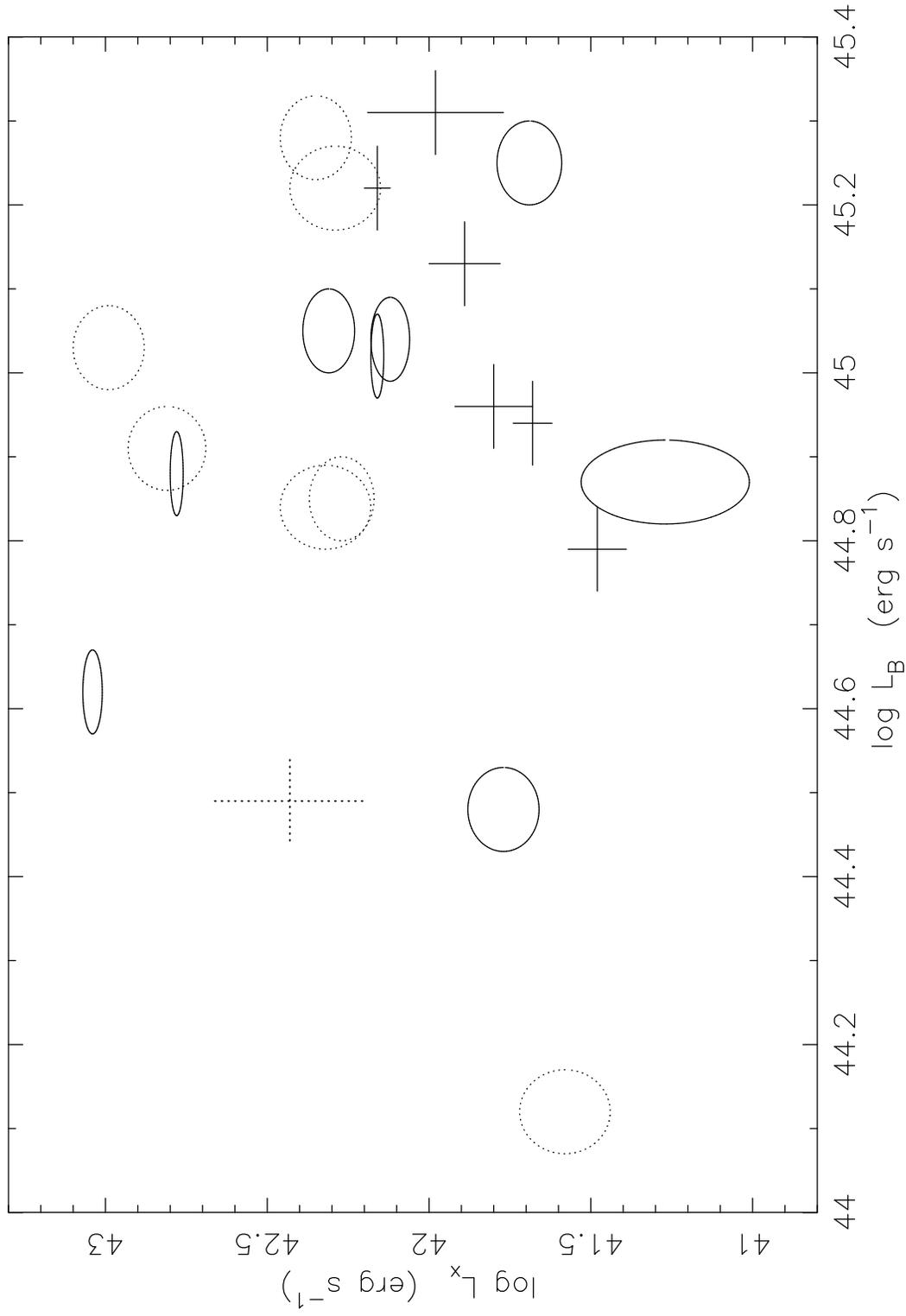
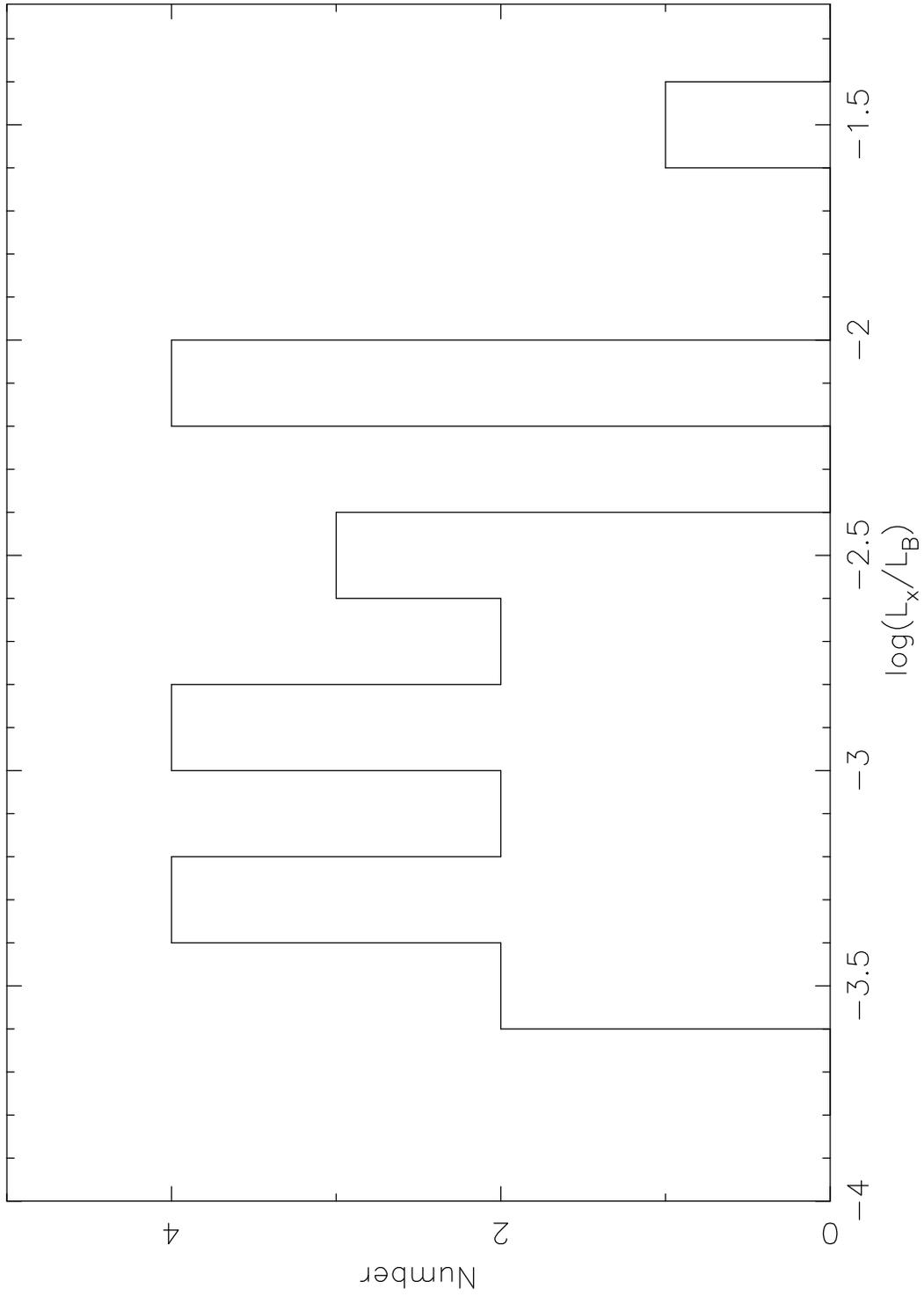
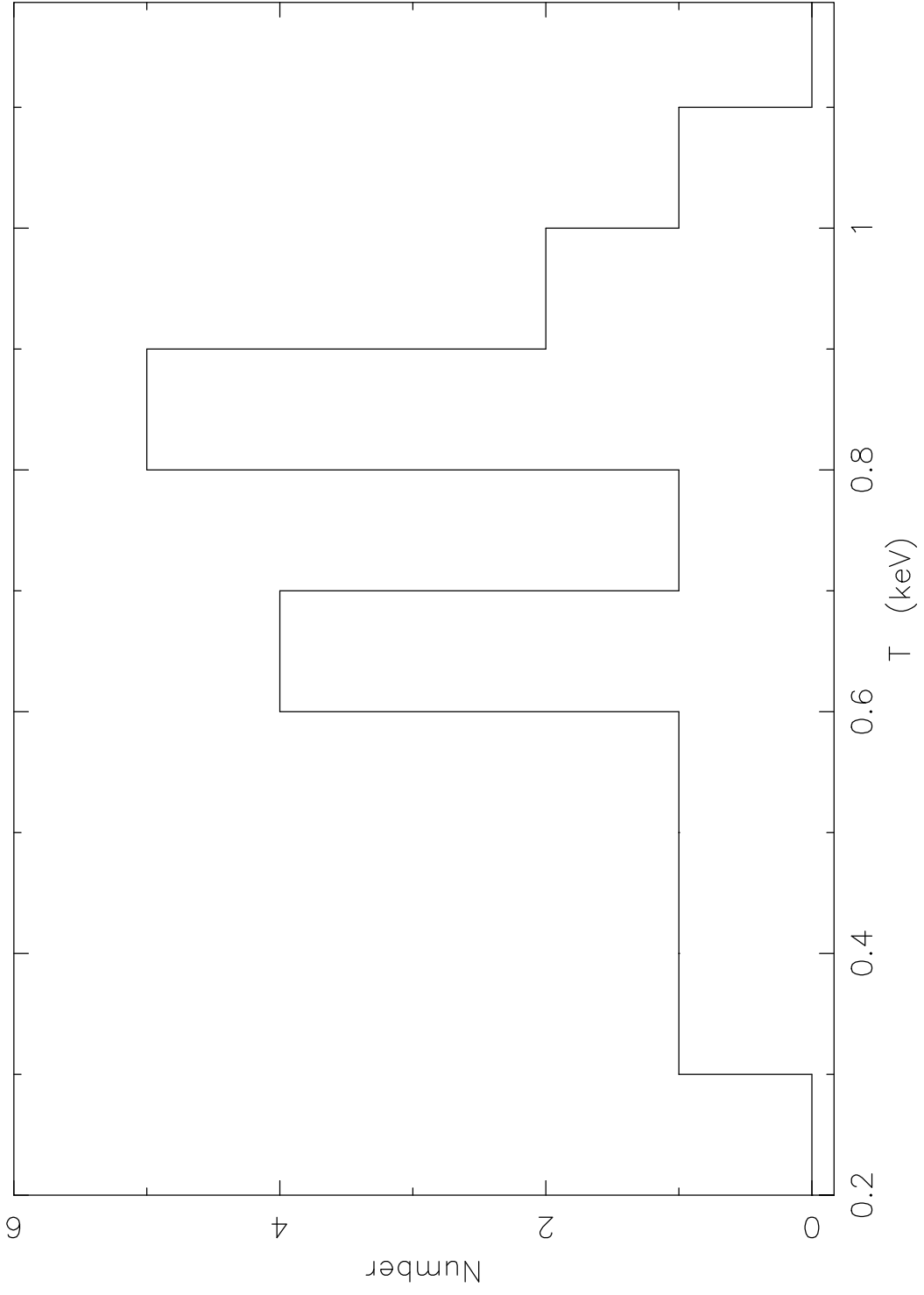
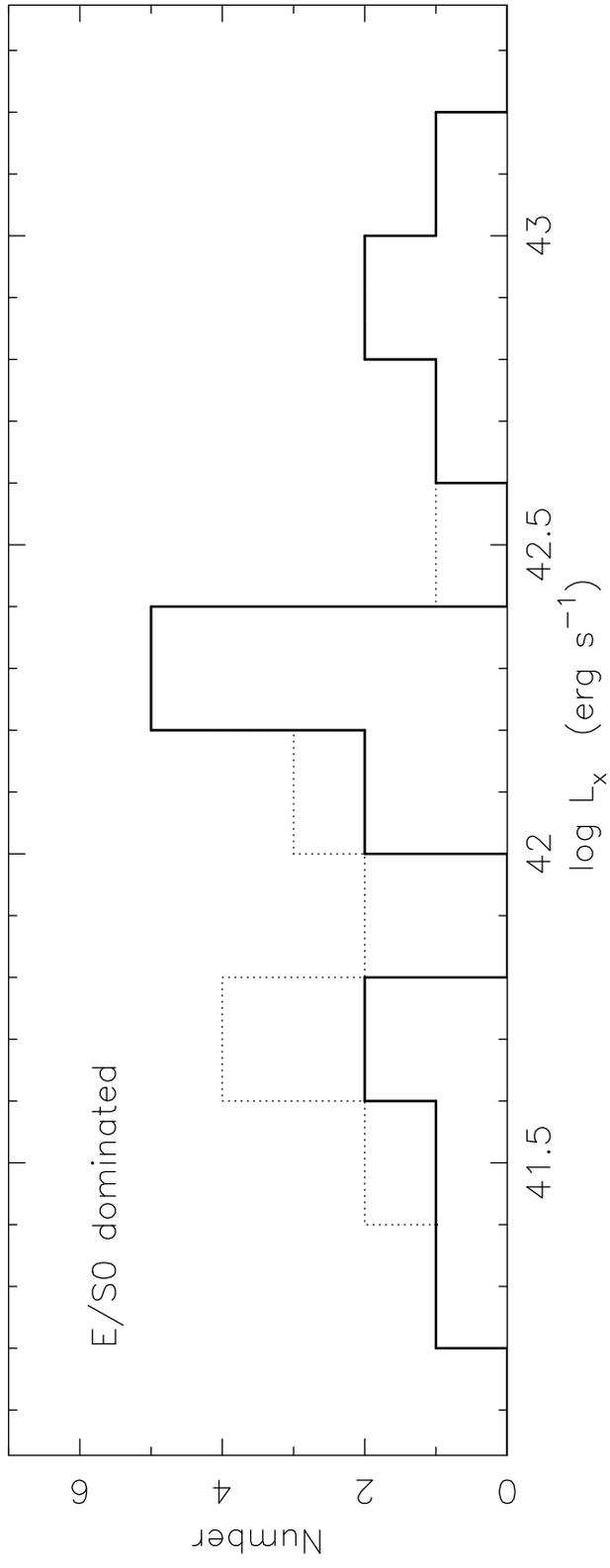
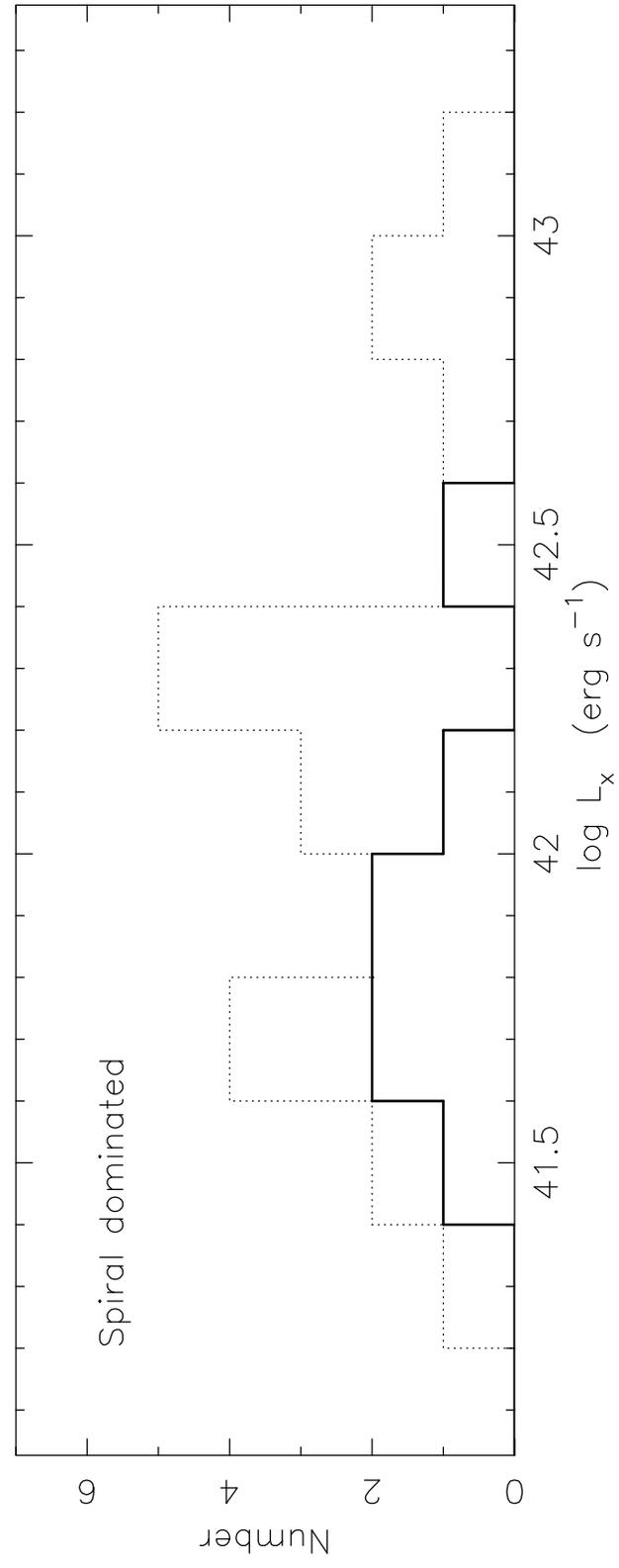
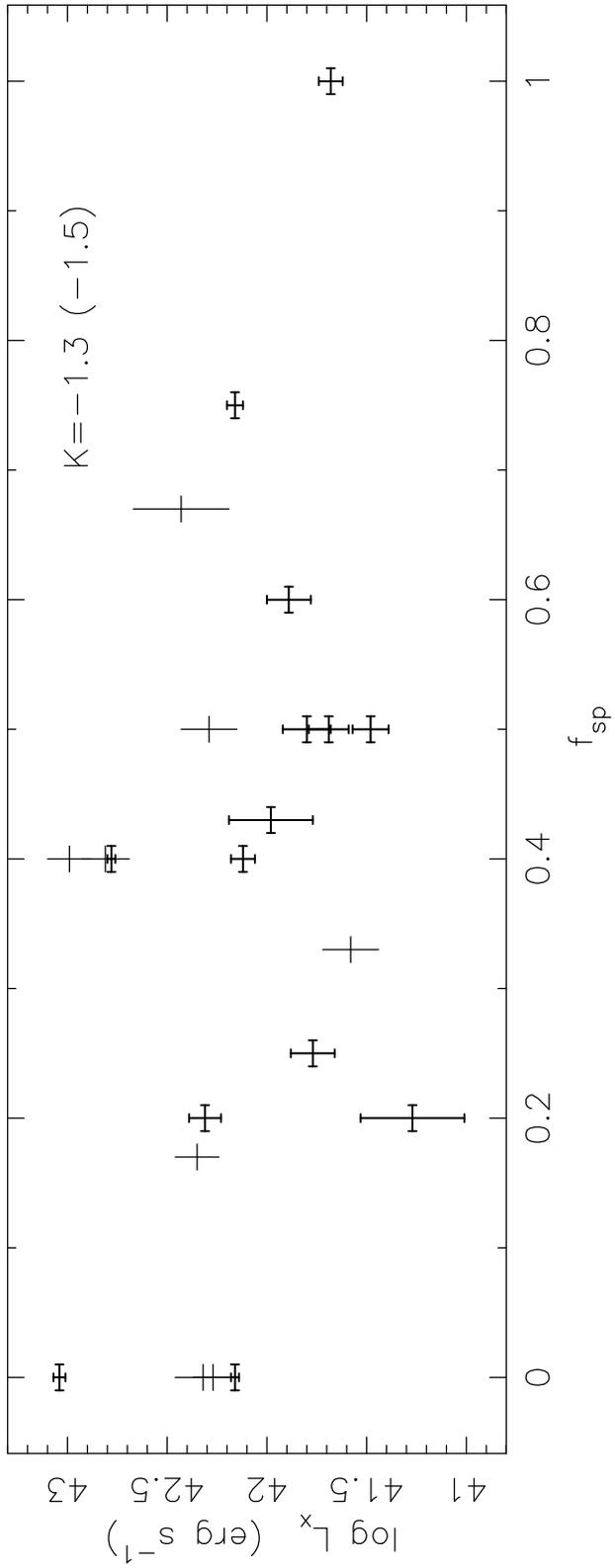
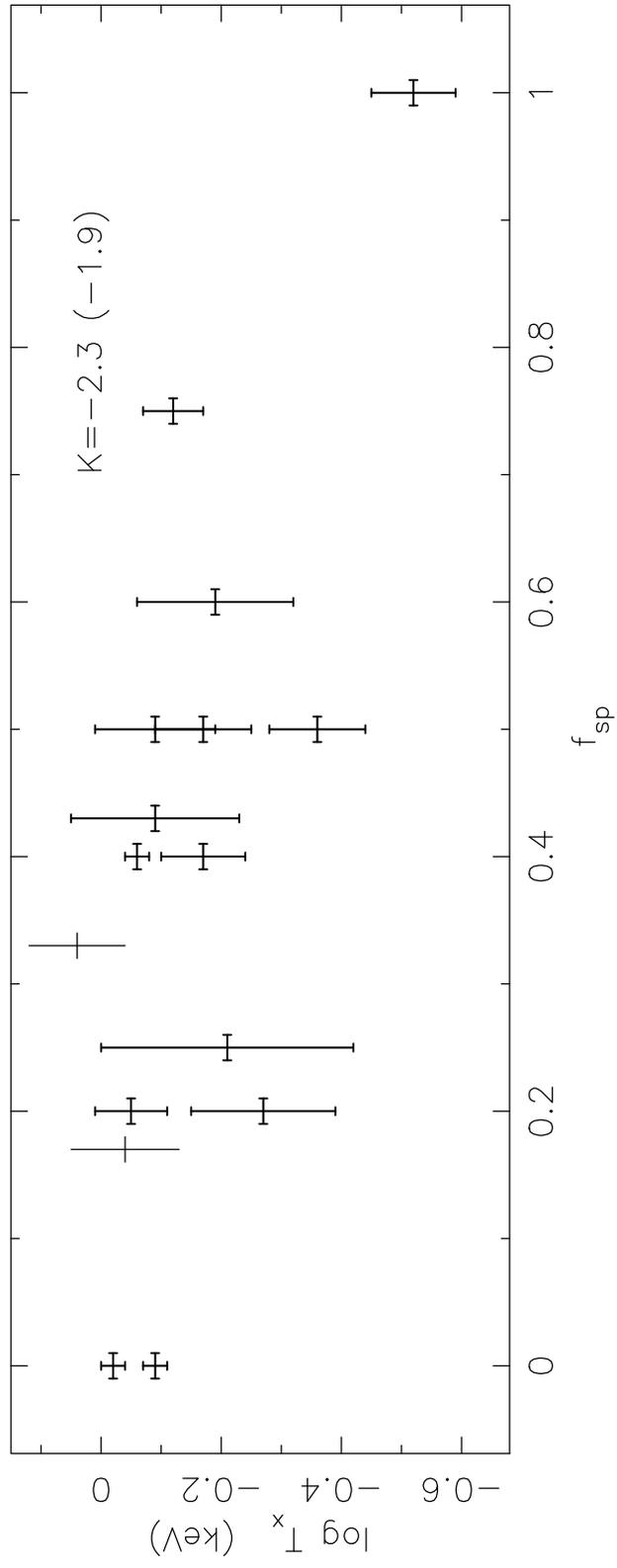
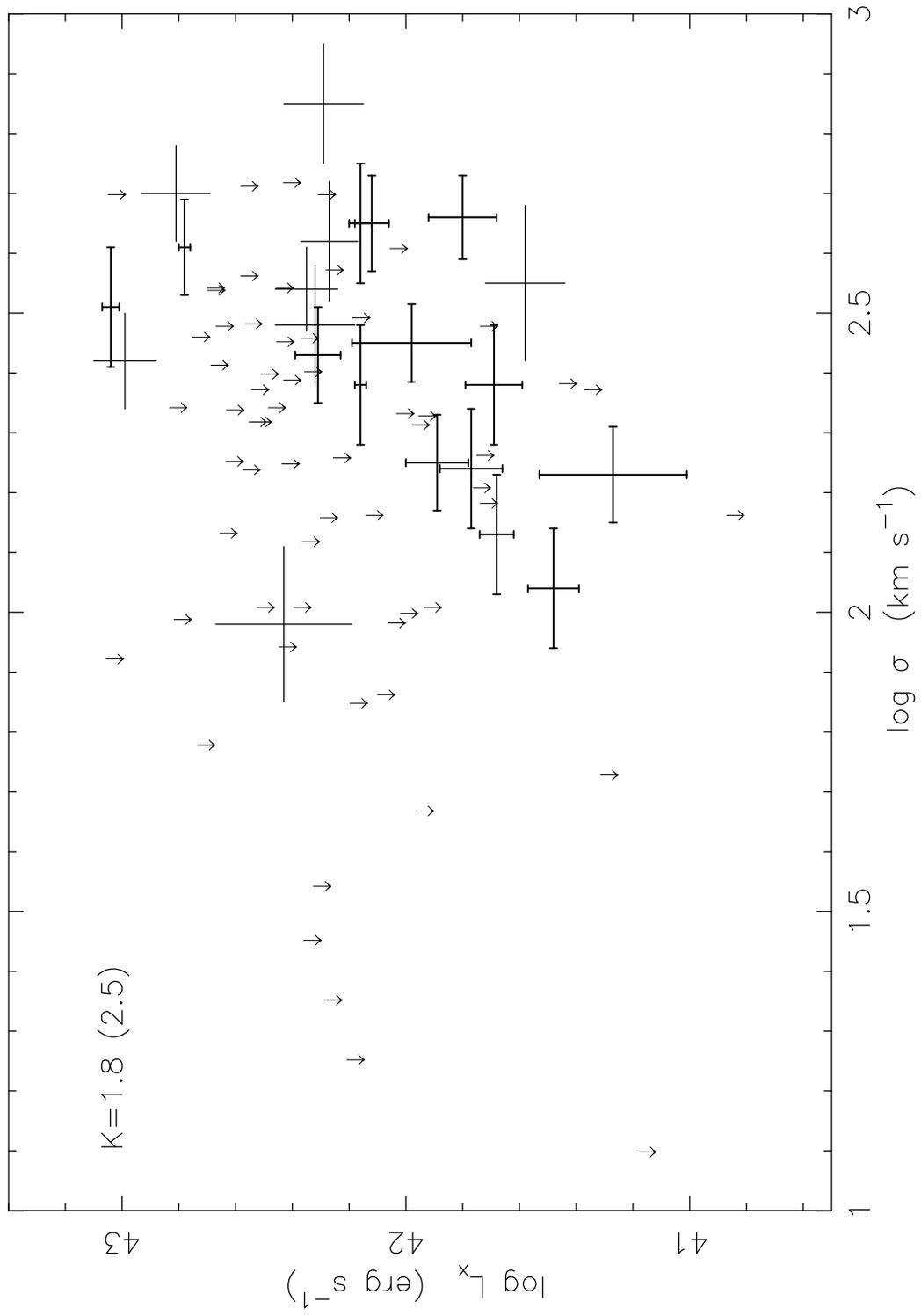
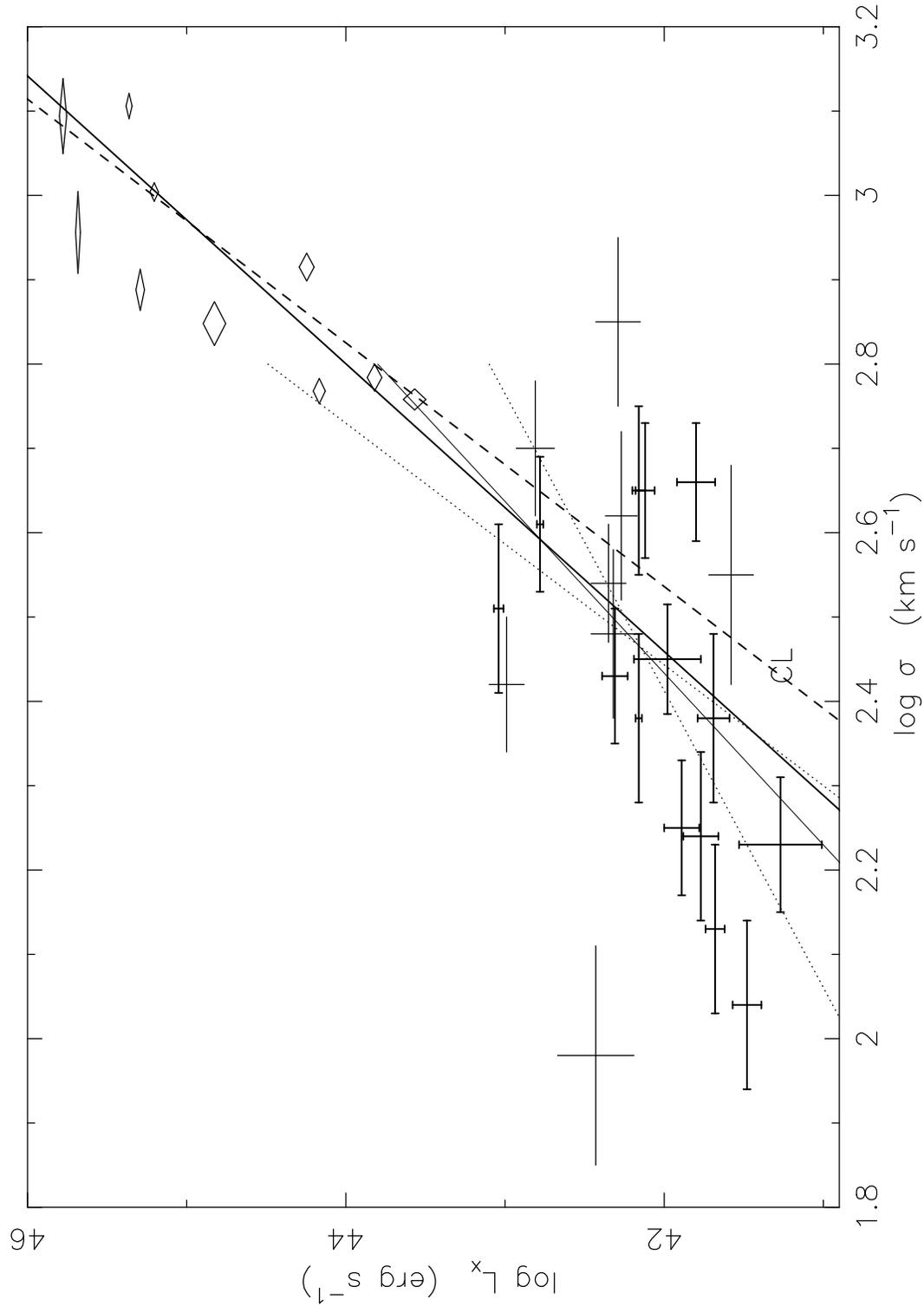
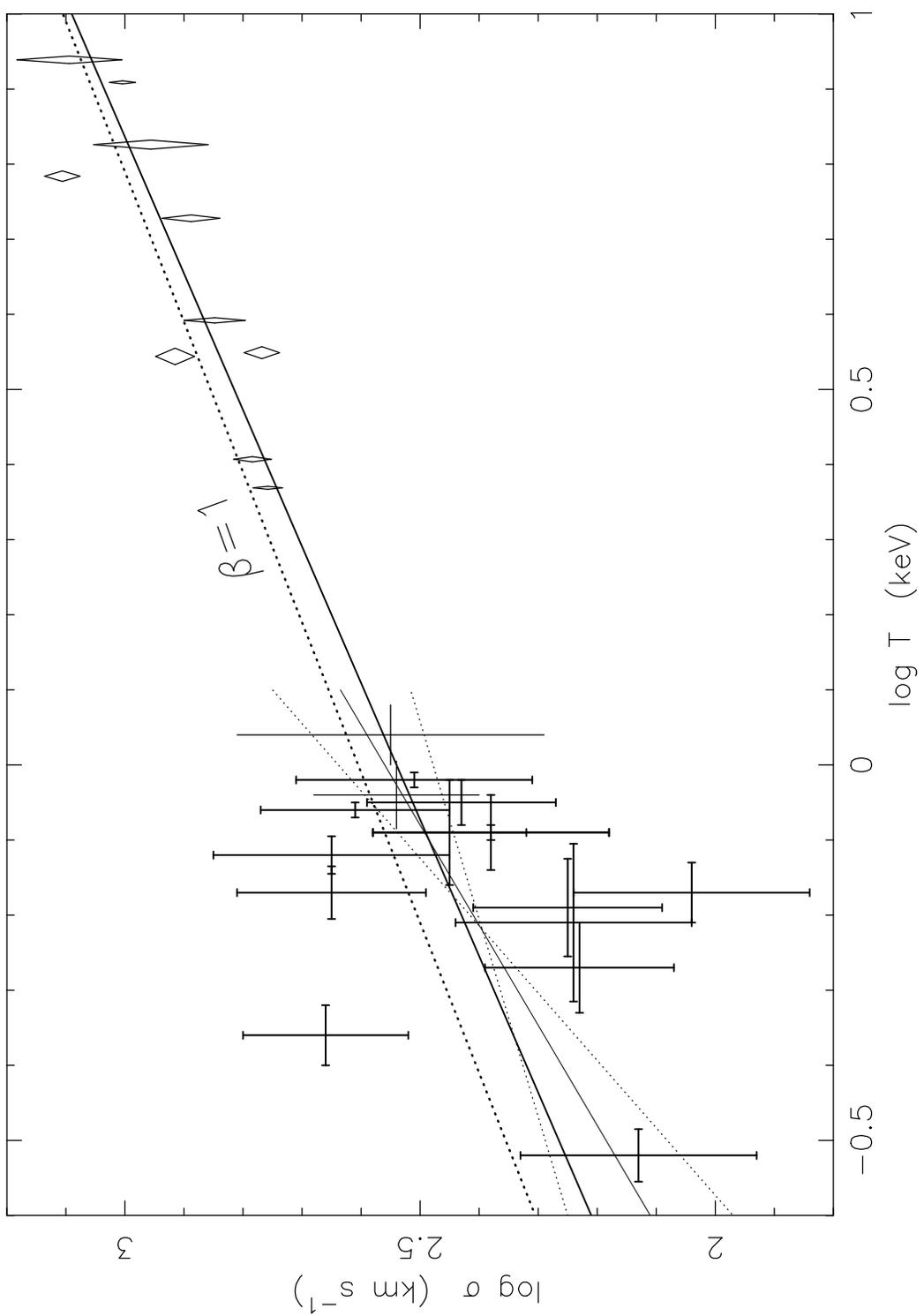
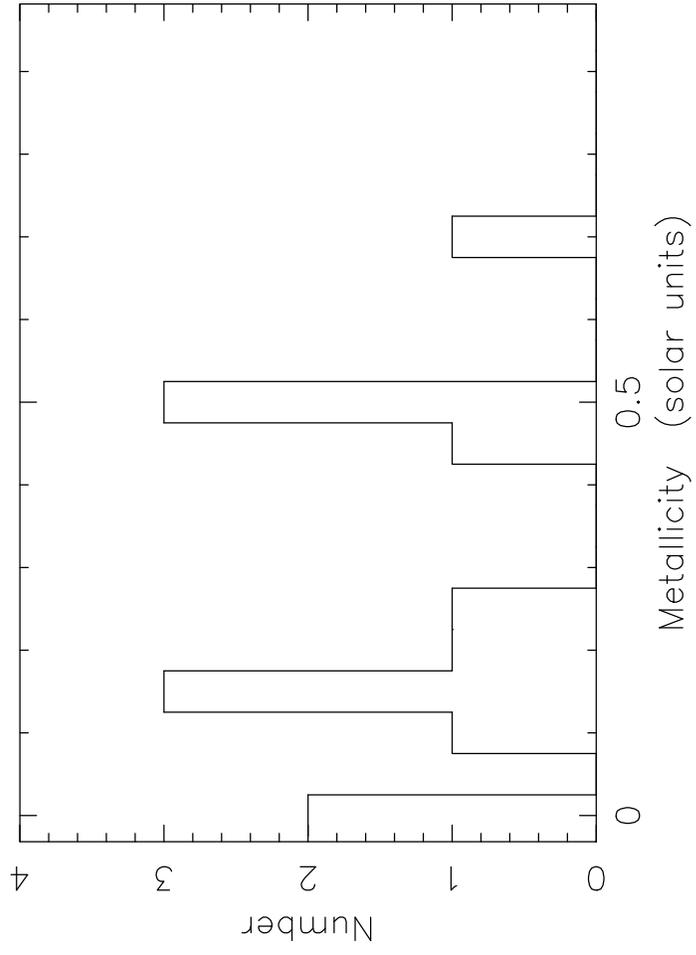
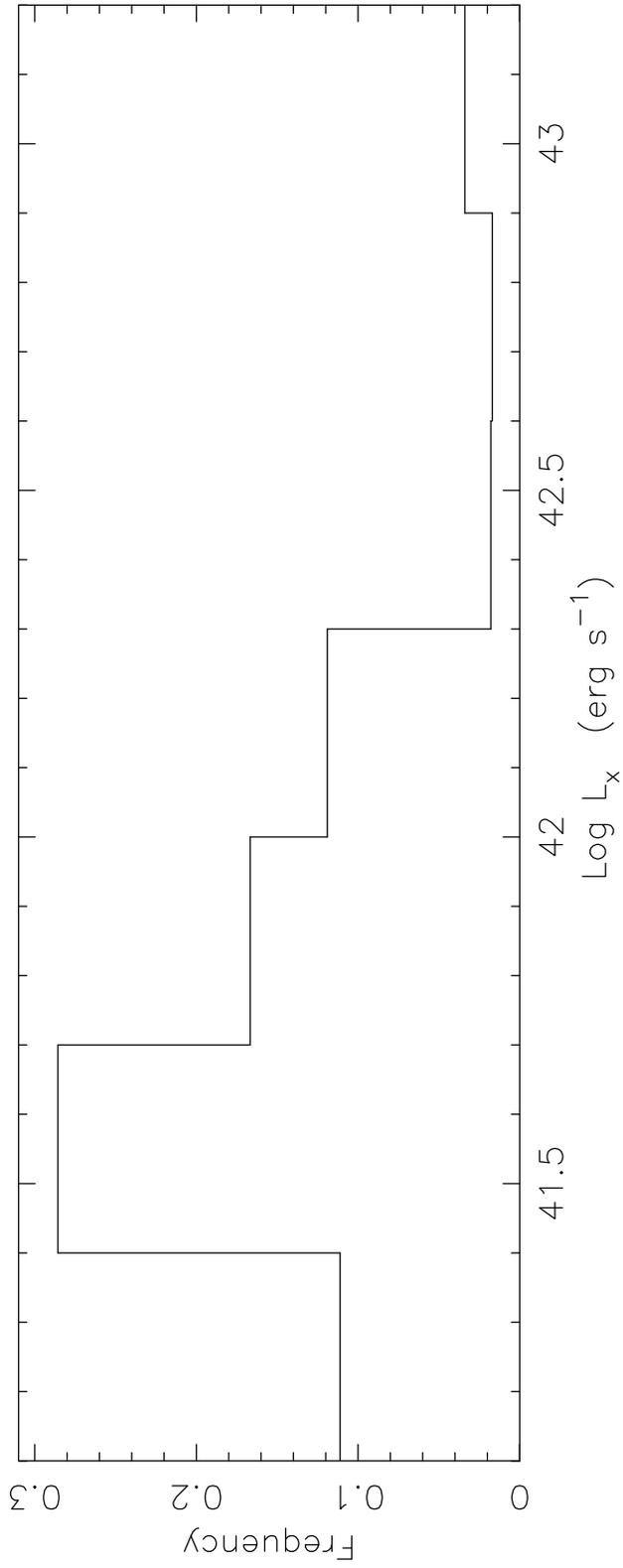
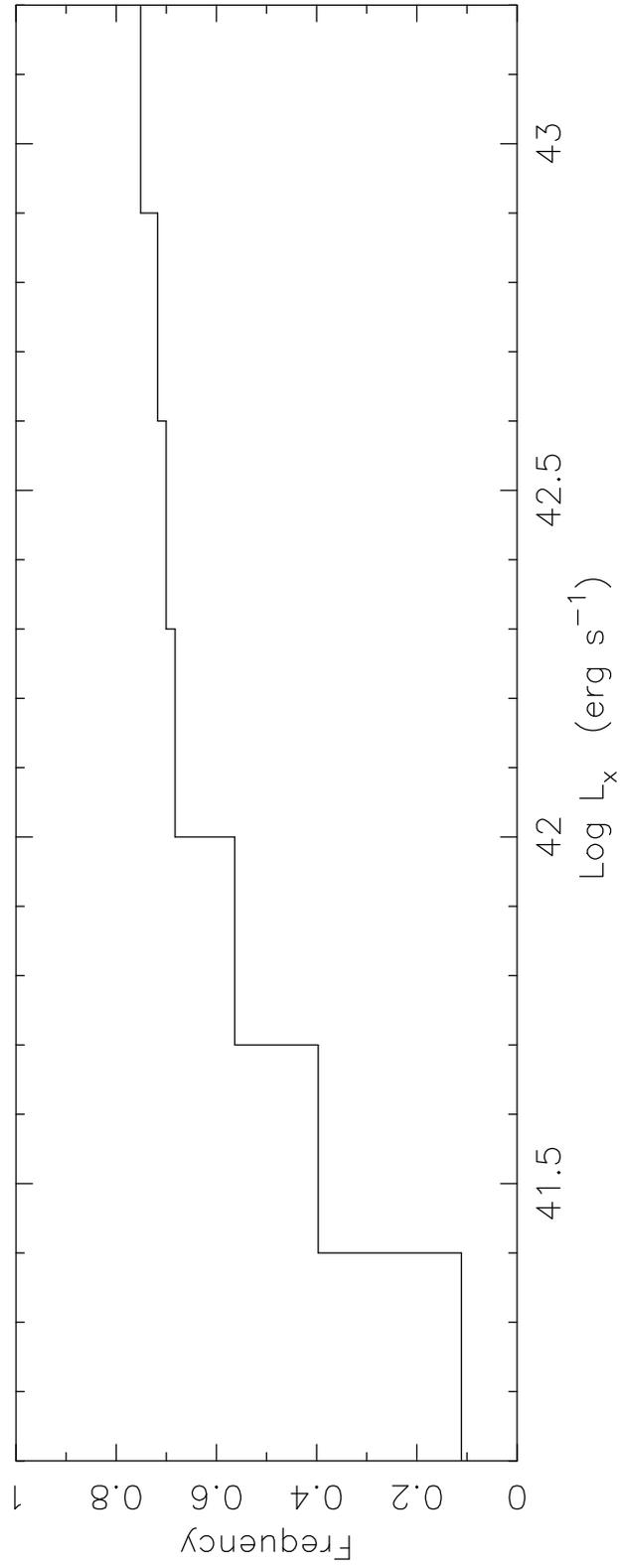
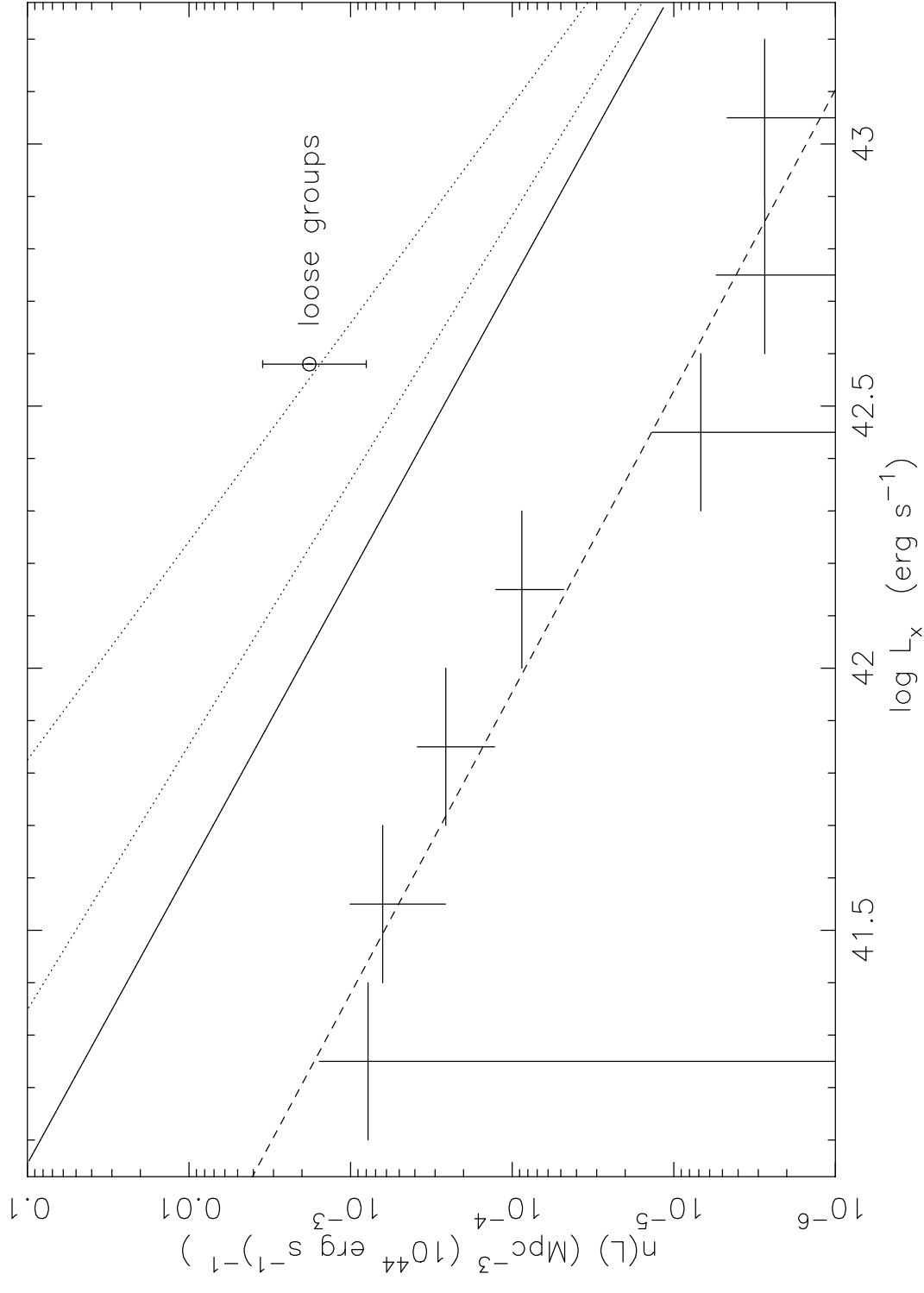
\end{document}